\documentclass[twocolumn,times]{aastex631}
\usepackage{amsmath}

\begin{document}

\title{Photometric and Spectroscopic Observations of GRB\,190106A: Emission from Reverse and Forward Shocks with Late-time Energy Injection}

\correspondingauthor{Dong Xu, Wei-Hua Lei}
\email{dxu@nao.cas.cn, leiwh@hust.edu.cn}

\author[0000-0002-9022-1928]{Zi-Pei Zhu}
\affiliation{Department of Astronomy, School of Physics, Huazhong University of Science and Technology, Wuhan, 430074, China}
\affiliation{Key Laboratory of Space Astronomy and Technology, National Astronomical Observatories, Chinese Academy of Sciences, Beijing, 100101, China}

\author[0000-0003-3257-9435]{Dong Xu}
\affiliation{Key Laboratory of Space Astronomy and Technology, National Astronomical Observatories, Chinese Academy of Sciences, Beijing, 100101, China}

\author[0000-0002-8149-8298]{Johan P. U. Fynbo}
\affiliation{The Cosmic Dawn Centre (DAWN), Niels Bohr Institute, University of Copenhagen, Lyngbyvej 2, 2100, Copenhagen, Denmark}

\author{Shao-Yu Fu}
\affiliation{Key Laboratory of Space Astronomy and Technology, National Astronomical Observatories, Chinese Academy of Sciences, Beijing, 100101, China}
\affiliation{School of Astronomy and Space Science, University of Chinese Academy of Sciences, Chinese Academy of Sciences, Beijing 100049, China}

\author{Xing Liu}
\affiliation{Key Laboratory of Space Astronomy and Technology, National Astronomical Observatories, Chinese Academy of Sciences, Beijing, 100101, China}
\affiliation{Key Laboratory of Cosmic Rays, Ministry of Education, Tibet University, Lhasa, Tibet 850000, China}

\author{Shuai-Qing Jiang}
\affiliation{Key Laboratory of Space Astronomy and Technology, National Astronomical Observatories, Chinese Academy of Sciences, Beijing, 100101, China}
\affiliation{School of Astronomy and Space Science, University of Chinese Academy of Sciences, Chinese Academy of Sciences, Beijing 100049, China}

\author[0000-0003-2957-2806]{Shuo Xiao}
\affiliation{Guizhou Provincial Key Laboratory of Radio Astronomy and Data Processing, Guizhou Normal University, Guiyang, 550001, People's Republic of China}
\affiliation{School of Physics and Electronic Science, Guizhou Normal University, Guiyang, 550001, People’s Republic of China}

\author[0000-0001-5553-4577]{Wei Xie}
\affiliation{Guizhou Provincial Key Laboratory of Radio Astronomy and Data Processing, Guizhou Normal University, Guiyang, 550001, People's Republic of China}

\author[0000-0002-5400-3261]{Yuan-Chuan Zou}
\affiliation{Department of Astronomy, School of Physics, Huazhong University of Science and Technology, Wuhan, 430074, China}

\author[0000-0002-3100-6558]{He Gao}
\affiliation{Department of Astronomy, Beijing Normal University, Beijing 100875, China}

\author[0000-0002-8028-0991]{Dieter Hartmann}
\affiliation{Department of Physics and Astronomy, Clemson University, Clemson, SC29634, USA}

\author[0000-0001-7717-5085]{Antonio de Ugarte Postigo}
\affiliation{Université Côte d'Azur, Observatoire de la Côte d'Azur, Artemis, CNRS, 06304 Nice, France}

\author[0000-0003-2902-3583]{David Alexander Kann}
\affiliation{Hessian Research Cluster ELEMENTS, Giersch Science Center, Max-von-Laue-Strasse 12, Goethe University Frankfurt, Campus Riedberg, 60438 Frankfurt am Main, Germany}

\author[0000-0003-3142-5020]{Massimo Della Valle}
\affiliation{Capodimonte Astronomical Observatory, INAF-Napoli, Salita Moiariello 16, Italy}
\affiliation{ Department of Physics, Ariel University, Ariel, Israel}
\affiliation{ICRANet, Piazza della Repubblica 10, I-65122 Pescara,Italy}

\author[0000-0002-9404-5650]{Pall Jakobsson}
\affiliation{Centre for Astrophysics and Cosmology, Science Institute, University of Iceland, Dunhagi 5, 107, Reykjavik, Iceland}

\author[0000-0003-3935-7018]{Tayabba Zafar}
\affiliation{Australian Astronomical Optics, Macquarie University, 105 Delhi Road, North Ryde, NSW 2113, Australia}
\affiliation{Macquarie University Research Centre for Astronomy, Astrophysics \& Astrophotonics, Sydney, NSW 2109, Australia}

\author{Valerio D'Elia}
\affiliation{ASI-Space Science Data Centre, Via del Politecnico snc I-00133, Rome, Italy}
\affiliation{INAF Osservatorio Astronomico di Roma, Via di Frascati 33, I-00040, Monteporzio Catone, Rome, Italy}

\author[0000-0002-9422-3437]{Li-Ping Xin}
\affiliation{Key Laboratory of Space Astronomy and Technology, National Astronomical Observatories, Chinese Academy of Sciences, Beijing, 100101, China}

\author{Jian-Yan Wei}
\affiliation{Key Laboratory of Space Astronomy and Technology, National Astronomical Observatories, Chinese Academy of Sciences, Beijing, 100101, China}
\affiliation{School of Astronomy and Space Science, University of Chinese Academy of Sciences, Chinese Academy of Sciences, Beijing 100049, China}

\author{Xing Gao}
\affiliation{Xinjiang Astronomical Observatory, Chinese Academy of Sciences, Urumqi,  Xinjiang 830011, China}

\author{Jin-Zhong Liu}
\affiliation{Xinjiang Astronomical Observatory, Chinese Academy of Sciences, Urumqi,  Xinjiang 830011, China}
\affiliation{Key Laboratory of Space Astronomy and Technology, National Astronomical Observatories, Chinese Academy of Sciences, Beijing, 100101, China}

\author{Tian-Hua Lu}
\affiliation{Key Laboratory of Space Astronomy and Technology, National Astronomical Observatories, Chinese Academy of Sciences, Beijing, 100101, China}
\affiliation{School of Astronomy and Space Science, University of Chinese Academy of Sciences, Chinese Academy of Sciences, Beijing 100049, China}

\author[0000-0003-3440-1526]{Wei-Hua Lei}
\affiliation{Department of Astronomy, School of Physics, Huazhong University of Science and Technology, Wuhan, 430074, China}

\begin{abstract}

Early optical observations of gamma-ray bursts can significantly contribute to the study of the central engine and physical processes therein. However, of the thousands observed so far, still only a few have data at optical wavelengths in the first minutes after the onset of the prompt emission. Here we report on GRB\,190106A, whose afterglow was observed in optical bands just 36 s after the {\em Swift}/BAT trigger, i.e., during the prompt emission phase. The early optical afterglow exhibits a bimodal structure followed by a normal decay, with a faster decay after $\sim \rm T_{0}+$1 day. We present optical photometric and spectroscopic observations of GRB\,190106A. We derive the redshift via metal absorption lines from Xinglong 2.16-m/BFOSC spectroscopic observations. From the BFOSC spectrum, we measure $z= 1.861\pm0.002$. The double-peak optical light curve is a significant feature predicted by the reverse-forward external shock model. The shallow decay followed by a normal decay in both the X-ray and optical light curves is well explained with the standard forward-shock model with late-time energy injection. Therefore, GRB\,190106A offers a case study for GRBs emission from both reverse and forward shocks. 

\end{abstract}

\keywords{\href{http://astrothesaurus.org/uat/629}{Gamma-ray bursts (629)}}

\section{Introduction} \label{sec:intro}
Since their discovery in the 1960s, the understanding of gamma-ray bursts (GRBs) has been greatly advanced by various space satellites and ground-based telescopes \citep[see][for a review]{2018pgrb.book.....Z}. Based on statistics of their prompt emission, GRBs are divided into two categories defined by their $T_{90}$ duration and spectral hardness, i.e., short bursts with hard spectra (also called type\,I bursts) with $T_{90}<2$ s and long bursts with soft spectra (also called type\,II bursts) with $T_{90}>2$ s \citep{1993ApJ...413L.101K,2009ApJ...703.1696Z}. The association of the short GRB\,170817A with the gravitational wave event GW170817 confirmed the longstanding prediction \citep{Eichler1989} that neutron star mergers can indeed produce short bursts \citep[]{Abbott+2017a, Abbott+2017b, Zhang+2018}. 
Long GRBs are generally believed to originate from massive star collapse \citep{2006ARA&A..44..507W}. GRB\,980425/SN 1998bw demonstrated that some stellar collapses produce long bursts and broad-lined Type Ic supernovae \citep{Galama+1999}. The comparison between long-duration GRBs and broad-lined SN rates suggests the ratio GRBs/CC-SNe is a few $\sim 10^{-3}$ \citep{2007ApJ...657L..73G}. 

Recent observations suggest that collapsing stars can also produce GRBs that last shorter than 2 s \citep{Ahumada+2021, Zhang+2021, Rossi+2021}, compact star mergers seem to be also able to produce long GRBs \citep{2022arXiv220410864R}, indicating that classification of GRBs is more complex than the simple temporal division. Since its launch in 2004, the {\it Neil\ Gehrels\ Swift\ Observatory} (\emph{Swift} hereafter) has detected more than 1700 GRBs with accurate position, enabling easy follow-up observations with ground-based telescopes \citep{2004ApJ...611.1005G}. The typical isotropic equivalent energy (since it is the energy equivalent to the one that the GRB should have if the emission were isotropic) of GRBs ranges between $10^{50}-10^{54}$ ergs within a few seconds to minutes \citep{2017ApJ...837..119A}. The jet break predicted by the collimated jet model has been observed in some GRBs, leading to the estimates of the true energy being $2-3$ orders of magnitude lower than the isotropic energy release \citep{2001ApJ...562L..55F}.

Previous studies have shown that the X-ray light curves of GRB afterglows typically have
several stages of power-law decay ($F \propto t^{-\alpha}$): a steep decay phase (typically $\alpha \sim 3$), a shallow decay phase also called a plateau (typically $\alpha \sim 0.5$), a normal decay phase (typically $\alpha \sim 1.2$), a post-jet-break phase (typically $\alpha \gtrsim 2$), with some bursts also accompanied by flares \citep{2006ApJ...642..354Z}. The typical optical light curve is an early rise followed by a normal decay, then sometimes followed by a re-brightening or a post-jet-break decay.
Some long bursts, when at low enough redshift to allow a search, are found to be accompanied by a supernova bump \citep{2004ApJ...609..952Z,Li+2012}. 

In the standard model of GRBs, the afterglow is believed to involve a relativistically expanding fireball. Considering a relativistic thin shell with energy $E_{\rm K}$, initial Lorentz factor $\Gamma_0$, opening angle $\theta_{\rm j}$, and a width in laboratory frame $\Delta_0$ expanding into the interstellar medium (ISM) with density $n$. The interaction between the shell and ISM is described by two shocks: 1) a forward shock (FS) propagating into the ISM. The shell first undergoes a coasting phase, where it moves at a nearly constant speed $\Gamma(t)\simeq \Gamma_0$. It starts to decelerate and evolves into the second stage when the mass of the ISM swept by the forward shock is about $1/\Gamma_0$ of the rest mass of the shell. The third phase is the post-jet-break phase when the $1/\Gamma$ cone becomes larger than the geometric cone defined by $\theta_{\rm j}$. Finally, the blastwave enters the Newtonian phase (its velocity is much smaller than the speed of light) when it has swept up the ISM with the total rest mass energy comparable to the energy of the shell. During all these three phases, electrons are believed to be accelerated at the forward shock front to a power-law distribution $N(\gamma_{\rm e}) \propto \gamma_{\rm e}^{-p_{\rm f}}$. A fraction $\epsilon_{\rm e,f}$ of the shock energy is distributed into electrons, and a fraction $\epsilon_{\rm B,f}$ is in the magnetic field generated behind the shock. Accounting for the radiative cooling and the continuous injection of new accelerated electrons at the shock front, one expects a broken power-law energy spectrum of them, which leads to a multi-segment broken power-law radiation spectrum at any epoch \citep[for a review]{Gao+2013}. The typical afterglow observations are successfully described by the synchrotron emission of electrons from the forward shocks. However, the information of the properties of the fireball is lost in the forward shock, especially when it has been decelerated by the ISM; 2) a reverse shock (RS) propagating into the shell, which occurs in the very early afterglow epoch. In the thin shell case, the reverse shock is too weak to decelerate the shell effectively. The Lorentz factor of the shocked shell material is almost constant while the shock propagates in the shell. The reverse shock crosses the shell at the fireball deceleration time. At this time, the reverse and forward shocked regions have the same Lorentz factor and internal energy density. However, the typical temperature of RS is lower since the mass density of the shell is higher. Consequently, the typical frequency of RS is lower, and RS may play a noticeable role in the low-frequency bands, e.g., optical or radio \citep{Meszaros1997}. After the RS crosses the shell, the Lorentz factor of the shocked shell may be assumed to have a general power-law decay behavior $\gamma_3 \propto r^{-g}$. The shell expands adiabatically in the shell's comoving frame. The emission from the RS is sensitive to the properties of the fireball. The observations of RS can thus provide important clues on the nature of the GRB source.

Catching very early (a few minutes after the trigger) optical emission from GRBs is not an easy task. There are only a few GRBs with optical observations during the prompt phase \citep[for a review]{2015AdAst2015E..13G}. \citet{2019A&A...628A..59O} provided a sample of 21 GRBs with at least one observation during the prompt emission phase, and several GRBs in the sample are well observed (eg., GRB\,070616, GRB\,081008, GRB\,121217A). Recently, \citet{2021arXiv210900010O} reported that they began to observe GRB\,210619B $\sim$ 28s after the \emph{Swift} trigger with D50 clear band during the prompt emission phase. However, our understanding of the earlier phases of GRBs continues to be quite incomplete compared to the later afterglow phases (occurring tens of minutes after the trigger). 

In this paper, we report a special burst, GRB\,190106A, for which optical observations began during the prompt emission phase, showing a double-peaked early optical light curve. We present optical photometric and spectroscopic observations of GRB\,190106A with the NEXT (Ningbo Bureau Of Education And Xinjiang Observatory Telescope) and the Xinglong 2.16-m Telescopes. With the Xinglong 2.16-m/BFOSC spectroscopic observation, we present the redshift and calculate the equivalent widths of absorption lines in the afterglow spectrum. Combining our data with the {\em Swift}/XRT data and other observations from GRB Coordinates Network (GCN) reports, we constrain the parameters in an external shock model of the afterglow.

Our observations and data reduction are described in Section \ref{Obs}. The spectroscopic data analysis and redshift measurements are presented in Section \ref{spec_analysis}.  The optical and X-ray afterglow data analysis and the modeling of the afterglow light curve are reported in Section \ref{temp_analysis}. We discuss possible models and present the conclusions in Section \ref{discussion} and Section \ref{conclusion}, respectively. A standard cosmology model is adopted with $H_{0}=67.3\ \rm{km\cdot s^{-1} Mpc^{-1}}$, $\Omega_{M}$=0.315, $\Omega_{\Lambda}$=0.685 \citep{Planck+2014}.

\section{Observations}\label{Obs}
GRB\,190106A triggered \emph{Neil Gehrels Swift Observatory} (short as \emph{Swift}) Burst Alert Telescope (BAT, \citealt{2005SSRv..120..143B}) at 13:34:44 UT on 2019 Jan 06 \citep{Sonbas+2019}. It is a long-duration burst with $T_{90}=76.8\pm2.4$ s.
Konus-\emph{Wind} also detected the burst with $T_{90}=77.1 ^{+1.3}_{-4.2}$ s  \citep{Tsvetkova+2019}. 
The MASTER-Amur robotic telescope quickly slewed to the burst position, starting observations 36 s after the BAT trigger. An optical transient (with a magnitude of 15.23 in the \emph{P}-band)  was discovered within the {\em Swift}/BAT error circle in the second 10 s exposure image \citep{Yurkov+2019,Lipunov+2019} located at equatorial coordinates (J2000.0): $RA=1^h 59^m 31.19^s$, $Dec.=+23^{\circ} 50^{'} 44.79^{''}$, consistent with the {\em Swift}/UVOT position $RA=01^h59^m31.18^s$, $Dec.\ =\ +23^{\circ}50^{'}44.0^{''}$ \citep{Kuin+2019}. XRT and UVOT started observations at 81.8 s and 90 s after the BAT trigger, respectively. 
The BAT data files were downloaded from the \emph{Swift} data archive\footnote{\url{https://www.swift.ac.uk/swift\_portal/}}. We extracted the BAT light curve and spectrum by the \emph{HEASoft} software package (v6.27.2). The 15-150 keV spectra are analyzed in \emph{XSPEC} (version 12.12) with power-law approximation over the time interval from 71 s to 81 s after the BAT trigger. We adopted the analysis results of the XRT repository products \citep{Evans+2007,Evans+2009} and downloaded the 0.3-10 keV unabsorbed light curve from the UK {\em Swift} Science Data Centre\footnote{\url{https://www.swift.ac.uk/xrt\_curves/}}.

\begin{figure*}[htp]
\center
\includegraphics[width=0.9\textwidth]{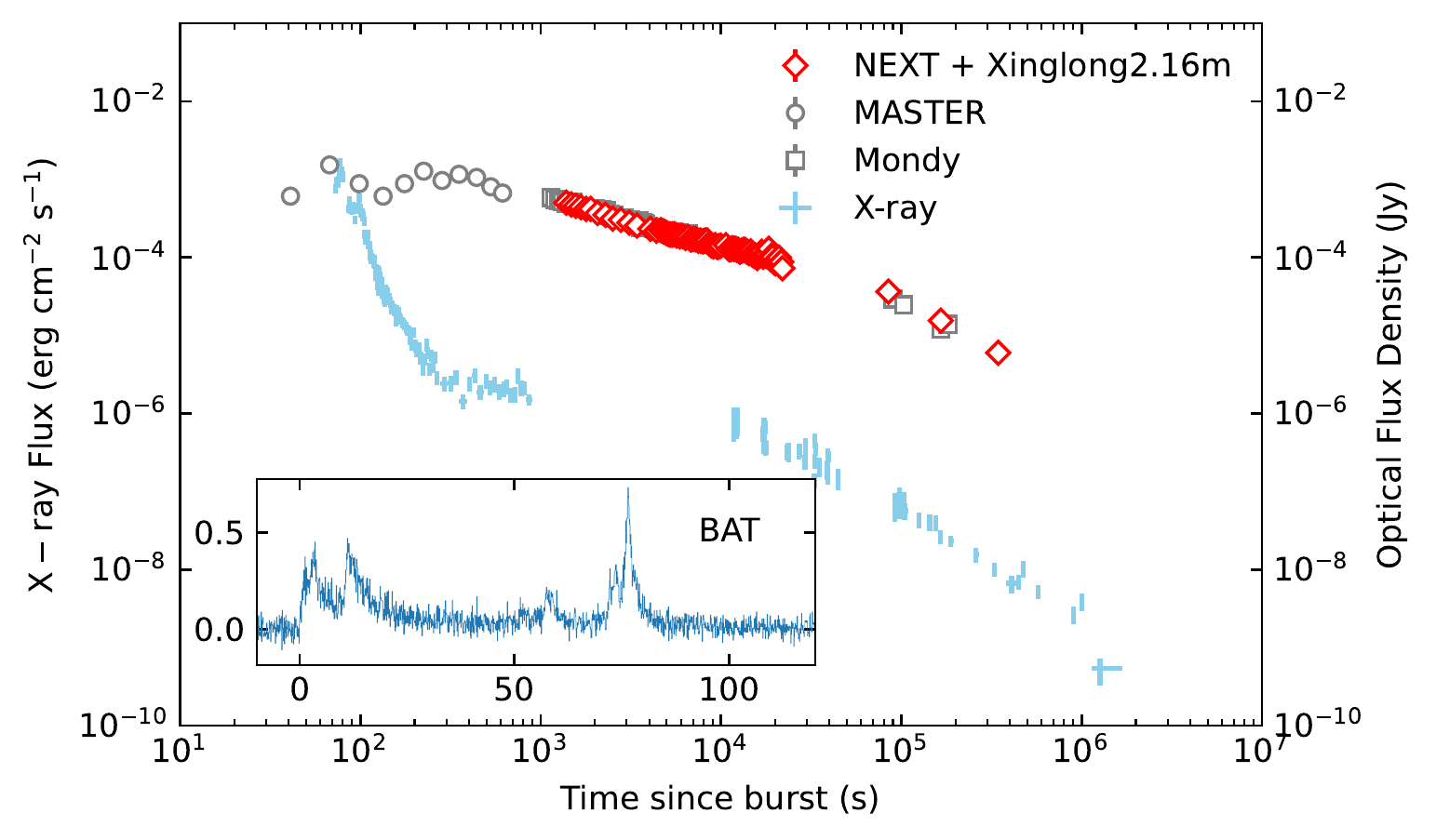}
\caption{The light curves of GRB\,190106A in $\gamma$-ray, X-ray and optical $R$-band. The X-ray light curve colored in cyan is 0.3-10 keV unabsorbed flux, obtained from UK {\em Swift} Science Data Centre \citep{Evans+2009}. The gray circles and gray squares are modified MASTER and Mondy photometric results, respectively. The red diamond points are our observations obtained by NEXT and Xinglong 2.16-m. Most of the error bars are smaller than the marker, and our photometric data are listed in the Table.~\ref{table:light curve}.}
\label{fig:light curve}
\end{figure*}

In our subsequent ground-based optical follow-up observations, the Xinglong 2.16-m obtained spectroscopic observations from which the redshift was determined \citep{Xu+2019}, also confirmed by GMG \citep{Mao+2019} and X-shooter \citep{Schady+2019}. We also collected observations from the GCNs reported by MASTER \citep{Lipunov+2019} and Sayan Observatory, Mondy \citep{Belkin+2019a, Belkin+2019b}. The MASTER data are converted to $R$-band by adding $\Delta R = 0.51$ mag following \citet{Xin+2018}. The corrected MASTER magnitude and Mondy optical observations are shown in Fig.~\ref{fig:light curve} in gray color.

\begin{figure*}[htp]
\center
\includegraphics[width=0.9\textwidth]{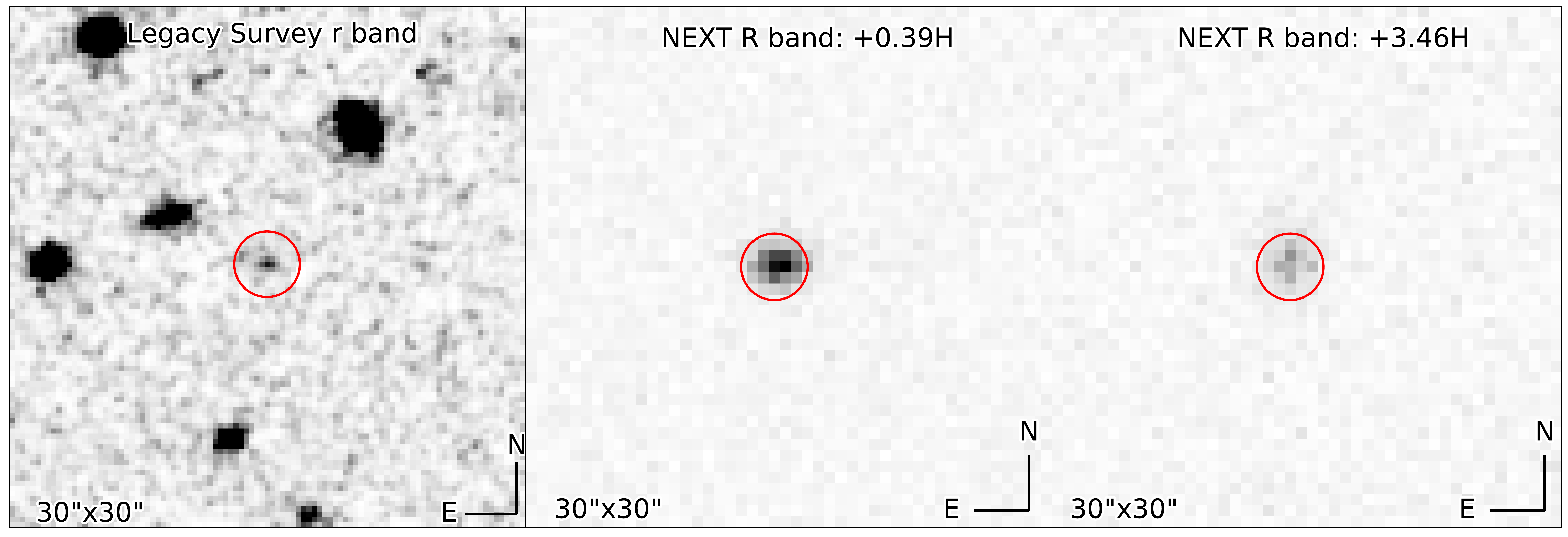}
\caption{The $R$-band position of GRB\,190106A within the field of view $30 ''\times 30''$. The location of the burst is circled in red. North is up and East is left. \textit{\bf Left panel}: the DESI Legacy Imaging Surveys $r$ band image \citep{2019AJ....157..168D}. A weak but obvious source with $r \sim 24.0$ mag is located at the GRB position, which could be the host galaxy of the burst. \textit{\bf Middle panel}: the first NEXT image, obtained 0.39 hours after the BAT trigger, shows a bright source consistent with the UVOT coordinates with $R=16.9$ mag. \textit{\bf Right panel}: the NEXT image obtained at 3.46 hours after the burst. The source in the image center is obviously fainter than the one in the middle panel, with a brightness of $R=$ 18.4 mag.}
\label{fig:location}
\end{figure*}

\subsection{NEXT optical observations}
NEXT is a fully automated telescope with a diameter of 60 centimeters, located at Nanshan, Xingming Observatory. NEXT is equipped with a E2V 230-42 back-illuminated CCD, and the CCD controller was made by the Finger Lakes Instrumentation (FLI). The size of the CCD is $2048 \times 2048$ pixels, with a pixel size of $15 \rm \mu m$. The pixel scale is $0.64  \rm ''/pixel$, and the field of view (FOV) is $22 ' \times 22 '$ according to the size of the CCD. The typical gain is $1.85 \rm e^{-}/ADU$, and typical readout noise is $13\ e^{-}$ with $500\ \rm kHz$ readout speed. The equipped filters were in the standard Johnson-Cousins system.

NEXT is connected to GCN/TAN alert system, but did not observe GRB\,190106A immediately as the dome was still closed at the very beginning of the event. NEXT began to obtain the first image at 13:57:21 UT, 22.6 minute after the BAT trigger, and the final one ended at 20:09:26 UT when the airmass was $\sim 9$. The NEXT image of the burst location is also shown in Fig.~\ref{fig:location}. We obtained 88 images, all in the $R_C$ filter. Five of them were discarded for poor quality. The observing center time and exposure time of each image used are presented in Table~\ref{table:light curve}.

The data reduction was carried out by standard processes in the {\em IRAF} packages \citep{Tody+1986}, including bias, flat and dark current corrections. The cosmic rays were also removed by the filtering described in \citet{Dokkum+2001}.  The final five useful images were combined using the $imcombine$ task in {\em IRAF}. The astrometry was calibrated using {\em Astrometry.net} \citep{Lang+2010}. The magnitudes were measured using $Source Extractor$ \citep{BertinArnouts+1996} using a circular aperture with a nine-pixel diameter. The photometric calibration was derived using the $Sloan\ Digital\ Sky\ Survey$ 14$\rm ^{th}$ data release \citep{Abolfathi+2018}, with flux/mag conversion of the SDSS system into the Johnson-Cousins system\footnote{\url{https://www.sdss.org/dr12/algorithms/sdssUBVRITransform/##Lupton}}. The light curve is shown in Fig.~\ref{fig:light curve} and the photometric results are presented in Table~\ref{table:light curve}.

\begin{figure*}[htp]
\center
\includegraphics[width=0.9\textwidth]{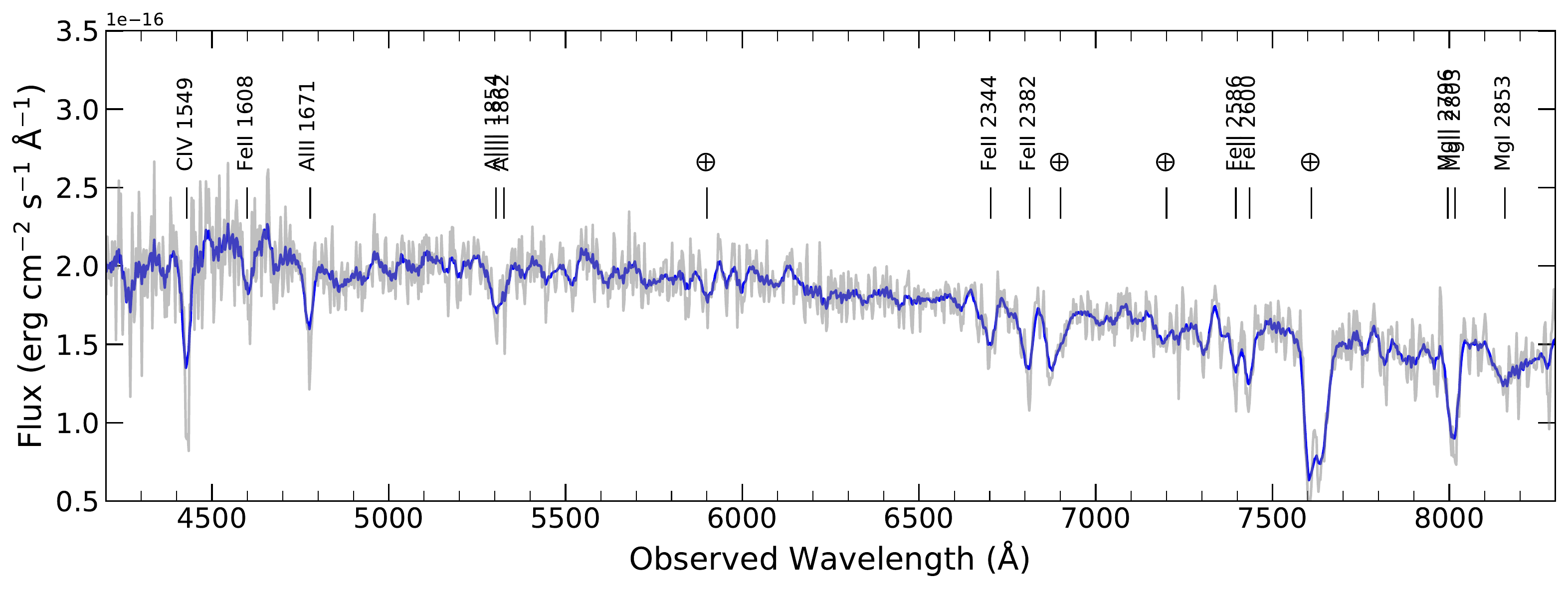}
\caption{The spectrum obtained with the Xinglong 2.16-m/BFOSC. Grey is the original spectrum and the blue spectrum has smoothed for display purposes. The identified metal absorption lines are indicated with vertical lines in the figure. Telluric features are marked with the telluric symbol.}
\label{fig:xl_spectrum}
\end{figure*}

\subsection{Xinglong 2.16-m Observations}
The spectroscopic observations were carried out with the National Astronomical Observatories, Chinese Academy of Sciences (NAOC) 2.16 m telescope \citep{Fan+2016} in Xinglong Observatory on 06 Jan 2019 at 14:02:39, $27.9$ minutes after the BAT trigger.  The optical spectrum was obtained with the Beijing Faint Object Spectrograph and Camera (BFOSC) equipped with a back-illuminated E2V55-30 AIMO CCD. The optical afterglow magnitude was $\sim16$ mag in \emph{R} band and the airmass was approximately 1.3 at the beginning of the spectroscopic observations. Due to a seeing of about $2\arcsec$ a slit-width of $1\farcs8$ was used. The slit was oriented in a south-north direction. The grating used was G4 and the order-sorter filter 385LP was also used, leading to a spectral coverage of 3800 to 9000 {\AA} at a resolution of about 10 {\AA}. A total of two spectra were obtained with exposure time of 3600 s and 2400 s, respectively.

The BFOSC spectroscopic data were reduced by the {\em IRAF} standard processes to obtain clean images, including bias subtraction, flat-field correction and cosmic-ray removal. With the using of ``NOAO'' package provided in {\em IRAF}, we extracted the one-dimensional spectrum of the burst, and calibrated the wavelength with arc-spectra from an iron-argon lamp. The flux-calibration was performed by the standard star HD+19445 \citep{Oke+1983} obtained with the same instrumental setup on the same night. After the processing above, the extracted spectrum is shown in Fig.~\ref{fig:xl_spectrum}.

\begin{figure}[htp]
\center
\includegraphics[width=0.45\textwidth]{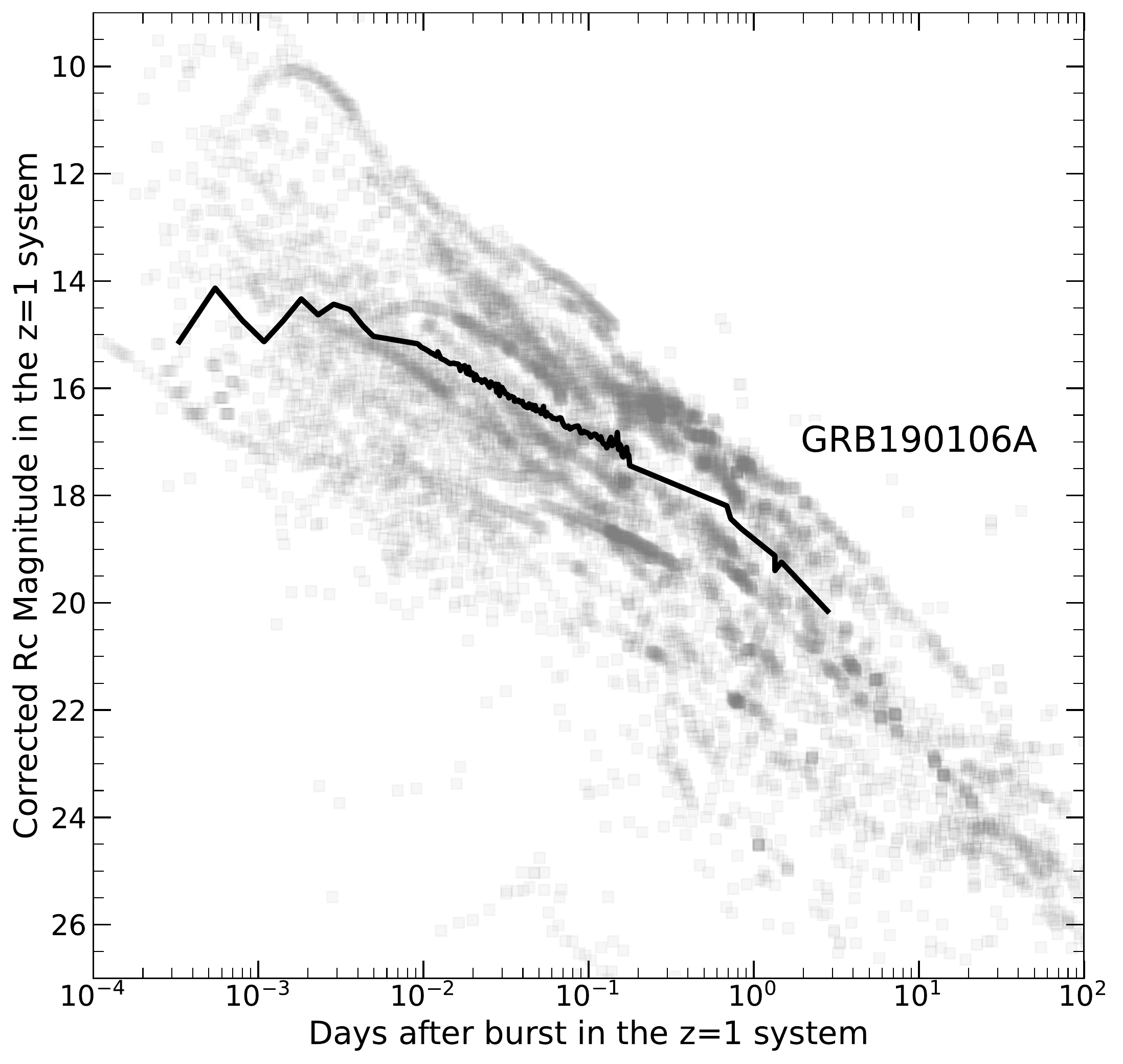}
\caption{Comparison of a large sample of GRB optical afterglow light curves shifted in time and flux to a common redshift of $z = 1$ following \citet{Kann2010}. The gray background are historical data  of other GRB light curves. GRB\,190106A is shown as a black solid line, which lies at the middle of the distribution in terms of luminosity.}
\label{fig:light curve comparision}
\end{figure}

The photometric observations were carried out on 7, 8 and 10 of Jan in the \emph{R} filter again using the BFOSC \citep{Fan+2016}. The exposure times were $5\times360$ s, $7\times360$ s, and $10\times360$ s, respectively. The data reduction and processing was the same as NEXT, except dark current, which was negligible. The subsequent magnitude measurements were also performed as described above. The photometric data are presented in Table~\ref{table:light curve} and shown in Fig.~\ref{fig:light curve}. A comparison with a sample of historical GRB optical light curves is shown in Fig.~\ref{fig:light curve comparision}. 

\section{Spectroscopic analysis}\label{spec_analysis}

As the blue end of the BFOSC spectrum has poor signal-to-noise ratio, our spectral analysis starts at $4300$~\AA. After excluding the A and B band absorption and sky emission lines, we identified the metal absorption lines detected at more than $3\sigma$ significance. We detected lines from \ion{C}{4}, \ion{Al}{2}, \ion{Al}{3}, \ion{Fe}{2}, and \ion{Mg}{2}. Line centroids were determined by fitting a Gaussian function to the line profiles. We obtain a redshift of $z=1.861\pm0.002$. The identified lines with the rest-frame equivalent widths ($\rm EW_{obs}$) are listed in Table~\ref{table:spectral lines} and also shown in Fig.~\ref{fig:xl_spectrum}.

\subsection{Line strength analysis}

\begin{figure*}[thp]
\center
\includegraphics[width=0.9\textwidth]{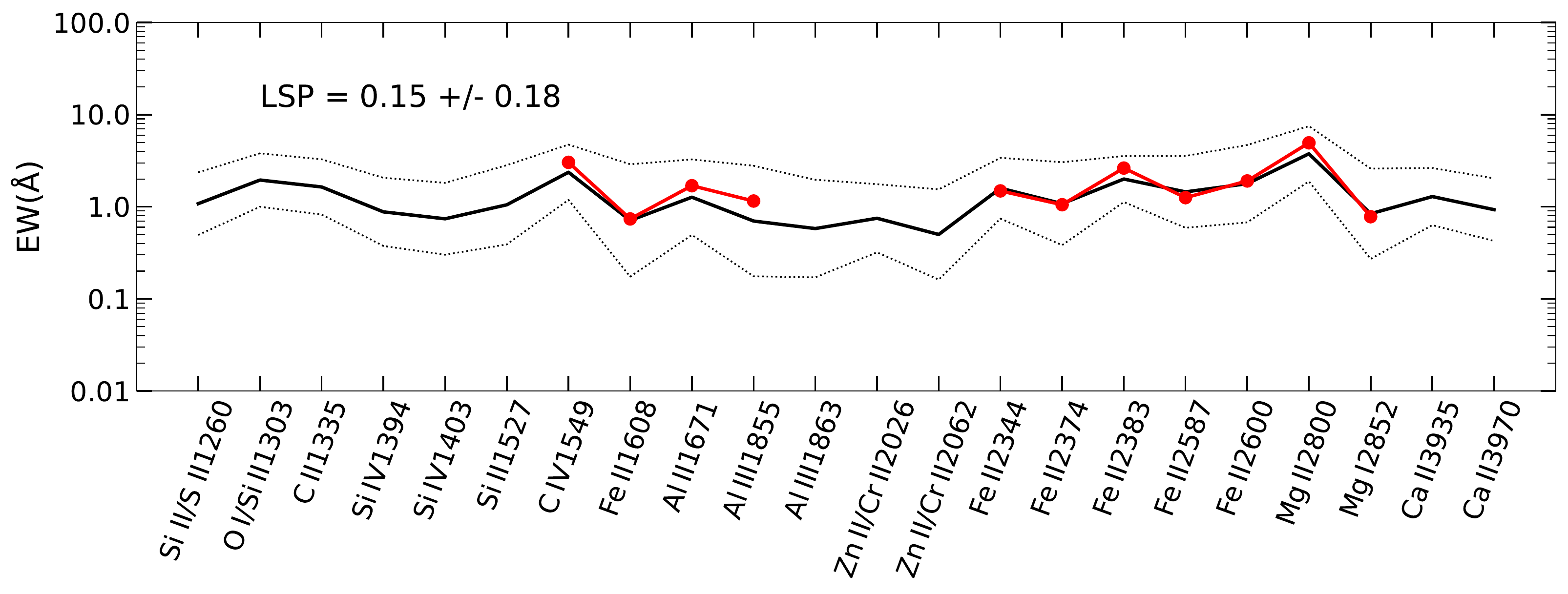}
\caption{The equivalent widths of GRB\,190106A obtained from BFOSC are colored in red. These can be compared to the average and standard deviation of strengths of a sample of GRB afterglows in black color, as described by \citet{Postigo+2012}.}
\label{fig:LSP}
\end{figure*}

We analyzed the strong features of the GRB host galaxy interstellar medium detected by BFOSC, following \citet{Postigo+2012}. The Line Strength Parameter (LSP) of the spectrum is $\rm LSP = 0.15 \pm 0.18$, which is very similar to the average value of the sample presented in the paper above. The line strength of the burst compared with the sample is shown in Fig.~\ref{fig:LSP}. 

\section{Optical and X-ray analysis and external shock Modeling}\label{temp_analysis}
\subsection{Temporal Analysis}

The \emph{R}-band, $\gamma$-ray and X-ray at 10 keV light curves are shown in Fig.~\ref{fig:light curve}. The optical light curve started at $T0+36$ s and ended at $T0+4.02$ d after the burst. According to the decay of the light curve, the optical and X-ray light curves can be divided into five and four stages, respectively. We fit the temporal indices $\alpha$ of the light curves in Fig.~\ref{fig:light curve} using $ F\ \propto\ t^{ - \alpha}$. 

The optical afterglow of the $R$-band light curve can be divided into five stages with different decay indices, which are listed in Table~\ref{tab:indices}, where the start and end times of different stages are also listed. The first stage is called \emph{fast rising}, from the beginning of observation to $\sim 67$ s after the burst. This is followed by a \emph{fast decay} until $\sim 1.3 \times 10^2$ s. In the third stage (\emph{slow rising}) it is still rising, the index has decreased by about 1 compared with the first stage. The fourth stage, \emph{normal decay}, it starts from $\sim 2.8 \times 10^2$ s and continues to $\sim 7.7\times10^4$ s with a decay index of 0.63. In the final stage, \emph{late decay}, the decay index increases to 1.29.

The light curve in X-rays seems simpler than in optical, and can be divided into the stages directly. The first stage is the tail of the prompt emission (\emph{steep decay}) with a decay index of $\sim4.2$, observations started at $T0+72.8$s. It is followed by a \emph{shallow decay} from $\sim2.6\times10^1$ s to $\sim1.6\times10^4$ s, with an index of 0.36. This stage can also be called a \emph{plateau phase}, resulting from the small index. In the final stage, which called \emph{late decay}, the decay index has increased to $\sim1.3$.

\begin{table}[htbp]
\begin{center}
\caption{List of X-ray and optical light curve decay indices.}
\label{tab:indices}
\begin{tabular}{ccccc}
\hline\hline
                 Band & Segment & Start (s)  & End (s) & $\alpha$  \\ 
    \hline
            X-ray  & steep decay  & - & $2.6\times10^2$ & $3.91 \pm 0.25$  \\
                   & shallow decay & $2.6\times10^2$ & $1.6\times10^4$ &  $0.36\pm0.03$ \\
                   & late decay & $1.6\times10^4$ & - &$1.31\pm0.03$ \\  
    \hline
        Optical   & fast rising  & - & $6.7\times10^1$  & $-1.88\pm0.57$ \\
                  & fast decay  & $6.7\times10^1$ & $1.3\times10^2$ & $1.35\pm0.32$  \\
                  & slow rising & $1.3\times10^2$ & $2.8\times10^2$ & $-0.58\pm0.17$ \\
                  & shallow decay & $2.8\times10^2$ & $7.7\times10^4$  & $0.63\pm0.01$ \\
                  & late decay & $7.7\times10^4$ & - & $1.29\pm 0.06$ \\ 
    \hline\hline
\end{tabular}
\end{center}
\end{table}

\begin{figure}[thp]
\center
\includegraphics[width=0.45\textwidth]{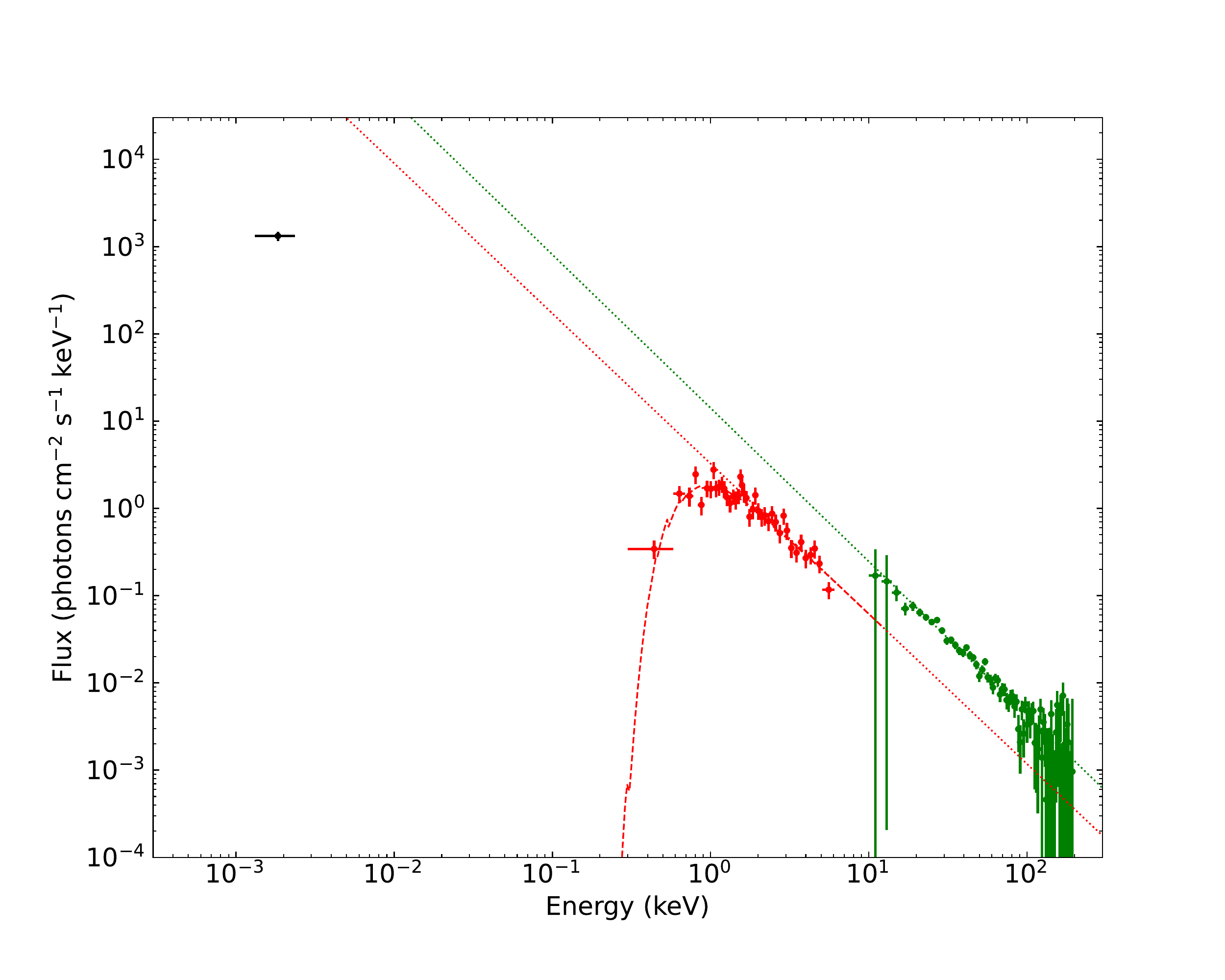}
\caption{The prompt phase SED combined with Optical (black), XRT (red) and BAT (green). The BAT time interval is from 71 s to 81 s, while the XRT is from 85 s to 95 s. The satellite was still slewing before 85 s, so the XRT data before 85 s are ignored. The optical point is converted from the MASTER observation and not included in the fitting. The data points of BAT and XRT were fitted with a single power-law function, and the spectral indices are $0.77\pm0.04$ and $0.72\pm0.10$, respectively.}
\label{fig:sed}
\end{figure}

The BAT ($15 - 350$ keV) count rates are also shown in the Fig.~\ref{fig:light curve}. The light curve can be simply divided into six episodes. The start, end and peak times of each episode are shown in Table~\ref{tab:bat}. We note that between $\sim$ 71 s and $\sim$ 81 s of the prompt emission, the optical light curve has a similar evolutionary behavior: they all peak around 70 s and then fade. However, the peak time is slightly different, i.e., $\sim$ 76 s for $\gamma-ray$, $\sim$ 76 s for X-ray, and  $\sim$ 68 s for optical. The two high energy bands are almost simultaneous, but the optics peak about 8 s earlier. In addition, the BAT and XRT spectrum of this period can be described using a single power-law $F_{\nu} \propto \nu^{-\beta}$, and $\beta_{\rm BAT} = 0.77\pm0.04$, $\beta_{\rm XRT} = 0.72\pm0.10$, which is shown in Fig.~\ref{fig:sed}. Although the UVOT optical data are not included in the spectral fitting, they are shown in the best-fit spectra in Fig.~\ref{fig:sed}. As we see, the optical data are much lower than the extrapolation of the best spectral fits for high energies. The SED containing optical and high energy (X-ray, $\gamma$-ray) can't be fitted by a single power-law. 

\begin{table}[htbp]
\begin{center}
\caption{The start, end and peak times of each episode. }
\label{tab:bat}
\begin{tabular}{ccccc}
\hline\hline
 Episode & Start(s)  & End(s) & Peak(s) & Duration(s)  \\ \hline
 Episode 1 & 0.2 & 9.2 & 2.8  & 9.0 \\
 Episode 2 & 9.2 & 23.5 & 11.0 & 14.3 \\
 Episode 3 & 50.6 & 57.1 & 52.1 &  6.5\\
 Episode 4 & 57.1 & 61.1 & 57.3 &  4.0\\
 Episode 5 & 71.0 & 75.2 & 73.3 &  4.2\\
 Episode 6 & 75.2 & 80.9 & 76.4 &  5.7\\
\hline\hline
\end{tabular}
\end{center}
\end{table}

\subsection{Afterglow SED analysis}

\begin{figure}[thp]
\center
\includegraphics[width=0.45\textwidth]{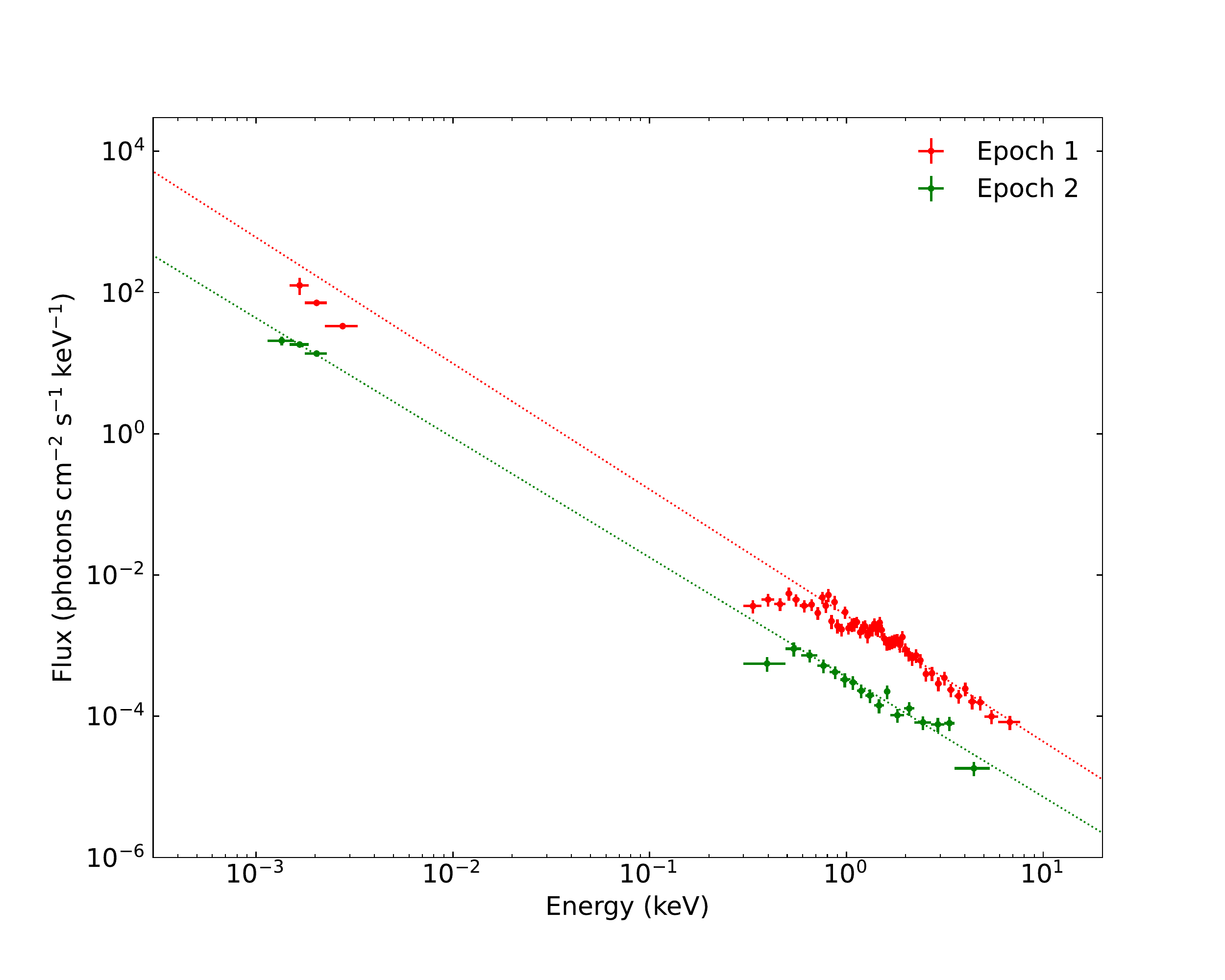}
\caption{The afterglow SED of GRB 190106A at 28.8 ks (Epoch 1, red lines) and 147.6 ks (Epoch 2, green lines) from optical to X-ray. The optical multi-band data are adopted from \citet{2019GCN.23660....1I} and \citet{2019GCN.23744....1D}. The dotted lines are the results of single power-law fit (dotted lines) for each epoch. The best fit photon indexes are $\Gamma=-1.78\pm0.03$ (Epoch 1) and $\Gamma=-1.79\pm0.01$ (Epoch 2).} 
\label{fig:afterglow_sed}
\end{figure}

To study the spectral energy properties of the afterglow, we select two epochs to construct its Spectral Energy Distribution (SED) with the best multi-band data: 28.8 ks (epoch 1) and 147.6 ks (epoch 2). The X-ray time sliced spectrum are obtained from the online repository\footnote{\url{https://www.swift.ac.uk/xrt\_spectra/}}, and the optical data are from \citet{2019GCN.23660....1I} and \citet{2019GCN.23744....1D}. We utilized the ``grpph'' tool of the \emph{Xspec} package to re-bin the X-ray data to guarantee that each bin contains at least 20 counts to improve the SNR. We fit the spectrum by the power law function from ``zdust*zphabs*phabs*powerlaw'' model in the \emph{Xspec} package, where ``zdust'' represents extinction by dust grains from the host galaxy of the burst, ``zphabs'' and ``phabs'' are neutral hydrogen photoelectric absorption of the host galaxy and  Galaxy, respectively. The redshift and the Galactic hydrogen column is fixed to 1.86 and $7.1\times10^{20}\ \rm cm^{-2}$ \citep{2016AA...594A.116H}, respectively. Small Magellanic Clouds extinction law is used as the host galaxy extinction model. The best fitted extinction E(B-V) of the host galaxy is 0.1 for epoch 1, which is also fixed in the fit of epoch 2. The best fitting result gives the photon index $\Gamma=-1.78\pm0.03$ and $\Gamma=-1.79\pm0.01$ for epoch 1 and 2, respectively. The results of these two epochs are shown in Fig.~\ref{fig:afterglow_sed}.

\subsection{Theoretical Interpretation}
The afterglow light curve of GRB\,190106A is unusual. There is an optical flash followed by a second bump in the early optical light curve, and then a shallow decay followed by a normal decay. The X-ray light curve shows a steeper decay at the end of the prompt emission, followed by a shallow decay, and finally a normal decay.

To understand the multi-band observational data of GRB\,190106A, we consider a relativistic thin shell with energy $E_{\rm K,iso}$, initial Lorentz factor $\Gamma_0$, opening angle $\theta_{\rm j}$, and a initial width in laboratory frame $\Delta_0$ expanding into the ISM with density $n$. The interaction between the shell and ISM is described by two shocks: an RS propagating into the shell and an FS propagating into the ISM \citep{Rees1992,Meszaros1997,Sari1998, Sari1999,Zou2005}.  There are four regions separated by the two shocks: Region 1, the ISM with density $n_1$; Region 2, the shocked ISM; Region 3, the shocked shell material; Region 4, the unshocked shell material with density $n_4$. First, the pair of shocks (FS and RS) propagating into the ISM and the shell, respectively. After the RS crosses the shell, the blastwave enters the deceleration phase. We now briefly describe these shocks separately.

The FS model we adopt was described in \citet{Lei2016}. The dynamical evolution of the shell are calculated numerically using a set of hydrodynamical equations \citep{Huang2000}
\begin{equation}
\frac{dR}{dt} =\beta c \Gamma (\Gamma+\sqrt{\Gamma^2 -1}),    
\end{equation}
\begin{equation}
\frac{dm}{dR} =2\pi R^2 (1-\cos\theta_{\rm j})n_1 m_{\rm p},   
\end{equation}
\begin{equation}
\frac{d\Gamma}{dm} = -\frac{\Gamma^2 -1}{M_{\rm ej} + 2\Gamma m} , 
\end{equation}
where $R$ and $t$ are the radius and time of the event in the burster frame, $m$ is the swept-up mass, $M_{\rm ej}=E_{\rm K, iso} (1-\cos\theta_{\rm j})/2(\Gamma_0-1)c^2$ is the ejecta mass, $m_{\rm p}$ is the proton mass, and $\beta=\sqrt{\Gamma^2-1}/\Gamma$. The last equation might be replaced with \citep{Geng2013}
\begin{equation}
\frac{d\Gamma}{dm} = -\frac{\Gamma^2 -1- \frac{1-\beta}{\beta c^3}L_{\rm inj} dR/dm}{M_{\rm ej} + 2\Gamma m} ,    
\end{equation}
when there is energy injection from GRB central engine. For $t_{\rm start}<t<t_{\rm end}$, the injected power is $L_{\rm inj}=L_{\rm inj}^0 (t/t_{\rm start})^{-q}$, where $L_{\rm inj}^0$ is the initial injection power, $q$ is the decay power law index, $t_{\rm start}$ and $t_{\rm end}$ are the start and end time for energy injection. By solving these equations with the initial conditions, one can find the evolution of $\Gamma(t)$ and $R(t)$. 

During the dynamical evolution of FS, electrons are believed to be accelerated at the shock front to a power-law distribution $N(\gamma_{\rm e}) \propto \gamma_{\rm e}^{-p_{\rm f}}$. Assuming a fraction $\epsilon_{\rm e,f}$ of the shock energy $e_2=4\Gamma^2 n_1 m_{\rm p} c^2$ is distributed into electrons, this defines the minimum injected electron Lorentz factor,
\begin{equation}
\gamma_{\rm m}=\frac{p-2}{p-1} \epsilon_{\rm e,f} (\Gamma-1)\frac{m_{\rm p}}{m_{\rm e}}    
\end{equation}
where $m_{\rm e}$ is electron mass. We also assume that a fraction $\epsilon_{\rm B,f}$ of the shock energy is in the magnetic field generated behind the shock. This gives the comoving magnetic field
\begin{equation}
B=(32\pi m_{\rm p} \epsilon_{\rm B,f} n_1)^{1/2} c.   
\end{equation}
The synchrotron power and characteristic frequency from electron with Lorentz factor $\gamma_{\rm e}$ are
\begin{equation}
P(\gamma_{\rm e})\simeq \frac{4}{3} \sigma_{\rm T} c \Gamma^2 \gamma_{\rm e}^2 \frac{B^2}{8\pi},   
\end{equation}
\begin{equation}
\nu(\gamma_{\rm e}) \simeq \Gamma \gamma_{\rm e}^2 \frac{q_{\rm e} B}{2\pi m_{\rm e}c},
\end{equation}
where $\sigma_{\rm T}$ is the Thomson cross-section, $q_{\rm e}$ is electron charge. The peak spectra power occurs at $\nu(\gamma_{\rm e})$
\begin{equation}
P_{\nu,{\rm max}} \simeq \frac{P(\gamma_{\rm e})}{\nu(\gamma_{\rm e})}=\frac{m_{\rm e}c^2 \sigma_{\rm T}}{3q_{\rm e}} \Gamma B.   
\end{equation}
By equating the lifetime of electron to the time $t$, one can define a critical electron Lorentz factor $\gamma_{\rm c}$
\begin{equation}
\gamma_{\rm c} = \frac{6\pi m_{\rm e}c}{\Gamma \sigma_{\rm T}B^2 t},
\end{equation}
the electron distribution shape should be modified for $\gamma_{\rm e} >\gamma_{\rm c}$ when cooling due to synchrotron radiation becomes significant. Accounting for the radiative cooling and the continuous injection of new accelerated electrons at the shock front, one expects a broken power-law energy spectrum of them, which leads to a multi-segment broken power-law radiation spectrum separated by three characteristic frequencies at any epoch \citep[see Equations 15-17 therein]{Gao+2013}.The first two characteristic frequencies $\nu_{\rm e}$ and $\nu_{\rm c}$ in the synchrotron spectrum are defined by the two electron Lorentz factors $\gamma_{\rm e}$ and $\gamma_{\rm c}$. The third characteristic frequency is the self-absorption frequency $\nu_{\rm a}$, below which the synchrotron photons are self-absorbed (self-absorption optical depth larger than unity). The maximum flux density is $F_{\nu,{\rm max,f}}=N_{\rm e,2} P_{\nu,{\rm max}}/4\pi D^2$, where $N_{\rm e,2}=4\pi R^3 n_1/3$ is the total number of electrons in Region 2 (shocked ISM) and $D$ is the distance of the source.

Besides the numerical calculation of FS as described above, we can also provide an analytical description of the main properties of the evolution and emission of FS. Generally, the evolution includes four phases. The first is a coasting phase, in which we have $\Gamma(t)\simeq \Gamma_0$. In the second phase, the shell starts to decelerate at the deceleration time 
\begin{equation}
t_{\rm dec}= \left( \frac{3E_{\rm K,iso}}{16\pi n_1 m_{\rm p} \Gamma_0^8 c^5}  \right)^{1/3}    
\end{equation}
when the mass $m$ of the ISM swept by FS is about $1/\Gamma_0$ of the rest mass in the ejecta $M_{\rm ej}$. After $t_{\rm dec}$, the shell approaches the \citet{BM1976} self-similar evolution $\Gamma(t)\simeq (17E_{\rm K,iso}/1024\pi n_1 m_{\rm p}c^5t^3)^{1/8}$ and $R(t)\simeq (17E_{\rm K,iso} t/4\pi n_1 m_{\rm p}c)^{1/4}$. Later, as the ejecta is decelerated to the post-jet-break phase at the time 
\begin{equation}
t_{\rm j}\simeq 0.6 {\rm day} \left(\frac{\theta_{\rm j}}{0.1 {\rm rad}}  \right)^{8/3}  \left(\frac{E_{\rm K,iso}}{10^{53} {\rm erg}} \right)^{1/3} n_1^{-1/3},
\end{equation}
when the $1/\Gamma$ cone becomes larger than $\theta_{\rm j}$. Finally, the blastwave enters the Newtonian phase when it has swept up the ISM with the total rest mass energy comparable to the energy of the ejecta. The dynamics is described by the well-known Sedov-Taylor solution. The synchrotron flux can be described by a series of power-law segments $F_\nu \propto t^{-\alpha} \nu^{-\beta}$ \citep{Sari1998,Gao+2013}. For example, in the second phase (deceleration phase), $\Gamma \propto t^{-3/8}$ and $R\propto t^{1/4}$ , one has the scalings for FS spectra parameters, i.e., $\nu_{\rm m}\propto t^{-3/2}$, $\nu_{\rm c} \propto t^{-1/2}$ and $F_{\nu, {\rm max,f} } \propto t^0$ \citep[see Equation 49 therein]{Gao+2013}. In the regime  $\nu_{\rm a}<\nu_{\rm m}<\nu<\nu_{\rm c}$, one has $F_{\nu,f} =F_{\nu, {\rm max, f}} (\nu/\nu_{\rm m})^{-\frac{p_{\rm f}-1}{2}} \propto t^{-\frac{3(p_{\rm f}-1)}{4}} \nu^{-\frac{p_{\rm f}-1}{2}}$. In the regime  $\nu_{\rm a}<\nu_{\rm m}<\nu_{\rm c}<\nu$, one has $F_{\nu,f} =F_{\nu, {\rm max, f}} (\nu_{\rm c}/\nu_{\rm m})^{-\frac{p_{\rm f}-1}{2}} (\nu/\nu_{\rm c})^{-\frac{p_{\rm f}}{2}} \propto t^{-\frac{3p_{\rm f}-2}{4}} \nu^{-\frac{p_{\rm f}}{2}}$ \citep[see Table 13 therein]{Gao+2013}. 

The RS model are described in \citet{Kobayashi2000}. In the thin shell case, the RS is Newtonian $\bar{\gamma}_{34}\sim 1$. The Lorentz factor of the shocked shell $\gamma_3$ is nearly constant during the shock crossing the shell. The shell begins to spread around $R_{\rm s}=\Gamma_0^2 \Delta_0$. The scalings before RS crossing time $t_{\rm dec}$ are \citep{Kobayashi2000}. 
\begin{align}
\gamma_3 \sim \Gamma_0,
\  n_3=(4\bar{\gamma}_{34}+3)n_4 \sim 7 n_1 \Gamma_0^2(t/t_{\rm dec})^{-3} , \nonumber \\
\ e_3\sim 4 \Gamma_0^2 n_1 m_{\rm p} c^2,
\ N_{\rm e, 3} \sim N_0 (t/t_{\rm dec})^{3/2},   
\end{align}
where $N_0=M_{\rm ej}/m_{\rm p}$ is the total number of electrons in the ejecta. 

We also assume that electrons are accelerated at the RS front to a power-law distribution $N(\gamma_{\rm e}) \propto \gamma_{\rm e}^{-p_{\rm r}}$, a fraction $\epsilon_{\rm e,r}$ of the shock energy $e_3$ is distributed into electrons and a fraction $\epsilon_{\rm B,f}$ to the magnetic field in Region 3. Follow the similar procedure as FS, we can obtain the scalings for parameters in the RS synchrotron spectrum \citep[see Equation 26 therein]{Gao+2013},
\begin{equation}
\nu_{\rm a} \propto t^{\frac{6p_{\rm r}-7}{p_r+4}}, \ \nu_{\rm m} \propto t^6, \ \nu_{\rm c} \propto t^{-2}, \ F_{\nu, {\rm max, r}} \propto t^{3/2}.   
\end{equation}
In the regime $\nu_{\rm m}<\nu_{\rm a}<\nu<\nu_{\rm c}$, one has $F_{\nu,r} =F_{\nu, {\rm max, r}} (\nu/\nu_{\rm m})^{-\frac{p_{\rm r}-1}{2}} \propto t^{3p_{\rm r}-3/2}$. 

After the RS crosses the shell, i.e.,$t>t_{\rm dec}$, the Lorentz factor of the shocked shell may be assumed to have a general power-law decay behavior $\gamma_3 \propto r^{-g}$. The shell expands adiabatically in the shell's comoving frame. The dynamical behavior in Region 3 are expressed with the scalings (with $g\sim 2$),
\begin{equation}
\gamma_3 \propto t^{-2/5}, \ n_3 \propto t^{-6/7}, \ e_3 \propto t^{-8/7}, \ N_{\rm e,3} =N_0.    
\end{equation}
 In the same way, we can obtain the scalings for parameters in the RS synchrotron spectrum for $t>t_{\rm dec}$ \citep[see Equation 31 therein]{Gao+2013},
\begin{align}
\nu_{\rm a} \propto t^{-102/175}, \ \nu_{\rm m} \propto t^{-54/35}, \ \nu_{\rm cut} \propto t^{-54/35}, \nonumber \\
F_{\nu, {\rm max, r}} \propto t^{-34/35},   
\end{align}
where $\nu_{\rm cut}$ is the cut-off frequency. After RS crossing, no new electrons are accelerated, $nu_{\rm c}$ will be replaced with $\nu_{\rm cut}$. In the regime $\nu_{\rm a}<\nu_{\rm m}<\nu<\nu_{\rm cut}$, one has $F_{\nu,r} =F_{\nu, {\rm max, r}} (\nu/\nu_{\rm m})^{-\frac{p_{\rm r}-1}{2}} \sim t^{-\frac{27p_{\rm r}+7}{35}}$. 

In this section, we will show that the double-peak optical light curve as well the X-ray observations of GRB 190106A can be well-explained by synchrotron emission from reverse and forward shocks with late-time energy injection. We adopt a numerical code described above for both FS and RS to model the multi-band observational data. The modeled X-ray and optical light curves corresponding to the above-mentioned parameters (i.e., kinetic energy $E_{\rm K,iso}$, initial Lorentz factor $\Gamma_0$ and opening angle of the relativistic shell, microphysics parameters for FS ($\epsilon_{\rm e,f}$,$\epsilon_{\rm B,f}$,$p_{\rm f}$) and RS ($\epsilon_{\rm e,r}$,$\epsilon_{\rm B,r}$,$p_{\rm r}$), and energy injection parameters ($L_{\rm inj}$,$q$,$t_{\rm start}$, $t_{\rm end}$) ) are displayed in Fig.~\ref{fig:model}. However, these model parameters still suffer degeneracy when fitting the optical and X-ray data. In this work, we do not attempt to fit the data across a large parameter space. We present a set of parameter values that interpret the data well, as shown in Table~\ref{tab:parametes}. The subscripts ``f '' and ``r'' denote the FS and RS emission, respectively.

\subsubsection{Early Double-peak Optical Light curve}
\citet{2003ApJ...595..950Z} pointed out that, depending on parameters, there are two types of early optical light curve for a fireball interacting with a constant-density ISM, that is, ``rebrightening'', in which a distinct reverse-shock peak (with $\alpha_{\rm rise} =-3p_{\rm r}+3/2$ and $\alpha_{\rm decay} =(27p_{\rm r}+7)/35$, see the description of RS model above and also Tables 4 and 5 in \citet{Gao+2013}) and forward-shock peak ($\nu_{\rm m,f}$ crossing peak) are detectable, and ``flattening'', in which the FS peak is buried beneath the RS peak. The early optical light curve during the time span $\sim40-1000$ s post-trigger shows a double-peak feature, which is clearly a ``rebrightening'' case. 

The first bump ($\sim40-133$ s) with rising temporal index $\alpha_{O} =-1.88\pm 0.57$ and decay index $\alpha_{O} =1.35\pm 0.32$ can be interpreted as the combinations of a reverse-shock peak (see the blue dashed line in Fig.~\ref{fig:model}) and forward-shock emission (see the blue dotted line in Fig.~\ref{fig:model}). We find the results with $p_{\rm r}=2.1$ can roughly explain such bump.

The first peak is also the RS crossing time (or deceleration time) $t_{\rm dec}$. We can estimate the initial Lorentz factor $\Gamma_0 \sim 300$, if we put $E_{\rm K,iso}=9\times 10^{52} \rm erg$ and $n_1=0.1 \rm cm^{-1}$ into Equation (11).

The second peak ($\sim133.8$ s) of the optical light curve occurs when the typical synchrotron frequency $\nu_{\rm m,f}$ crosses the observed frequency \citep{Sari1998}. The slow-rising index $\alpha_O =-0.58\pm 0.17$ of the second optical bump ($\sim133-1000$ s) is consistent with emission below $\nu_{\rm m}$ in a slow-cooling scenario, i.e., $F_{\nu,{\rm f}} \propto t^{1/2} \nu^{1/3}$ \citep[see Table 13 therein]{Gao+2013}. For such FS peak ($\nu_{\rm m,f}$ crossing peak), one has $F_{\nu,{\rm peak,f}} = F_{\nu,{\rm max,f}} \propto \epsilon_{\rm B,f}^{1/2} E_{\rm K, iso}  n_1^{1/2}$ \citep[see Equation 49 therein]{Gao+2013}. The values of jet isotropic kinetic energy $E_{\rm K, iso}$, ISM density $n_1$ and the microphysics parameter $\epsilon_{\rm B,f}$ are chosen to fit this peak.

\subsubsection{Shallow Decay Phase}
GRB\,190106A shows quite a long shallow decay in both X-rays (following the sharp decay) and the optical (following the second bump). In a slow-cooling regime of the forward shock, we expect $F_\nu\propto \nu^{-(p_{\rm f}-1)/2} t^{-3(p_{\rm f}-1)/4}$ for $\nu_{\rm m}<\nu<\nu_{\rm c}$, which corresponds to $\alpha = 3(p_{\rm f}-1)/4 \simeq 0.8$  using $p_{\rm f}=2.07$  \citep[see Table 13 therein]{Gao+2013} . Our observed values $\alpha_O=0.63\pm 0.01$ and $\alpha_X=0.36\pm 0.03$ are shallower than those expected. This phase can be understood within the standard external-shock model with nearly constant energy injection $L_{\rm inj} \propto t^{-q}$  (see Fig.~\ref{fig:model}, and \citealt{2012MNRAS.422.2044X} for a similar case). In Table~\ref{tab:parametes}, we present the parameter values that interpret the data well, i.e., the index $q\simeq 0.05$, initial injection power $L_{\rm inj}^0 \simeq 4.5\times 10^{49} {\rm erg \ s^{-1}}$, and starting time $t_{\rm start}>300/(1+z)$ s. The energy injection shuts off at $t_{\rm end} \sim 1500/(1+z)$ s.

There are two popular models for the energy injection, i.e., the spin-down of a magnetar and the fall-back accretion onto a stellar mass black hole (BH). First, we consider a fast-spinning magnetar as the central engine. The characteristic spin-down luminosity $L_0$ and the characteristic spin-down timescale $\tau$ are related to the magnetar initial parameters as \citep{Zhang2001}:
\begin{equation}
L_0 = 1.0\times 10^{49} \rm{erg\ s^{-1}} \left(B_{p,15}^2 P_{0,-3}^{-4}R_6^6\right),
\end{equation}
\begin{equation}
\tau = 2.05\times 10^3 \rm{s} \left(I_{45}B_{\rm{p},15}^{-2}P_{0,-3}^2 R_6^{-6}\right).
\end{equation}
For the NS equation of state (EoS), we adopt EoS GM1 (the radius of the magnetar $R=12.05 ~\rm km$ and the rotational inertia $I=3.33 \times 10^{45}~ \rm g\ cm^{-2}$) as suggested by the recent studies with GRB data \citep{lv2015,Gao2016}. Inserting $L_0=L_{\rm inj}=4.5\times 10^{49}\ {\rm erg \ s^{-1}}$ and $\tau\simeq 1500/(1+z)$ into the above two equations, one can thus infer a magnetar initial period of $P_0 \sim 1.7$ ms and a magnetic field $B_{\rm p} \sim 3.5 \times 10^{15} $ G.

For a black hole central engine model, the hyper-accreting BH system can launch a relativistic jet via neutrino-antineutrino annihilation \citep{1999ApJ...518..356P, 2001ApJ...557..949N, 2002ApJ...579..706D, 2004MNRAS.355..950J, 2006ApJ...643L..87G, 2007ApJ...657..383C, 2007ApJ...661.1025L, 2015ApJS..218...12L, 2009ApJ...700.1970L, 2016ApJ...833..129X} or Blandford-\.{Z}najek mechanism (hereafter BZ; \citealt{1977MNRAS.179..433B, 2000JKPS...36..188L, 2000ApJ...534L.197L, 2005ChJAS...5..279L, 2013ApJ...765..125L}).
The neutrino annihilation mechanism is too ``dirty'' to account for a GRB jet \citep{2013ApJ...765..125L, 2017ApJ...838..143X}. For this reason, we suppose that the energy injection is dominated by the BZ mechanism. The BZ power can be rewritten as a function of mass accretion rate as \citep{2013ApJ...765..125L}
\begin{equation}
L_{\rm BZ}=9.3 \times 10^{53} \frac{a_\bullet^2 \dot{m}  F(a_\bullet)}{(1+\sqrt{1-a_\bullet^2})^2} \ {\rm erg \ s^{-1}} ,
\label{eq:EB}
\end{equation}
where $\dot{m} \equiv \dot{M}/(M_\sun\ \rm s^{-1})$ is the dimensionless accretion rate, $a_\bullet=J_\bullet c/(GM_\bullet^2)$ is the spin parameter of the BH, $F(a_{\bullet})=[(1+q_a^2)/q^2][(q_a+1/q_a) \arctan q_a-1]$ and $q_a= a_{\bullet} /(1+\sqrt{1-a^2_{\bullet}})$. Assuming $a_\bullet=0.9$, the peak accretion rate is $\dot{M}\simeq 1.4\times 10^{-4} M_\sun\ \rm s^{-1}$.

As we can see, both central engine models can give rise to the energy injection required. However, a fall-back accreting BH tends to produce a giant bump in GRB afterglow light curves \citep{Wu2013}. The plateau phase is a natural expectation from a magnetar central engine \citep{Zhang2001,lv2015}. The small injection index $q\simeq 0.05$ (as shown in Table~\ref{tab:parametes}) favors the magnetar model.

\subsubsection{Late Decay Phase}

As shown in Table \ref{tab:indices}, just after the shallow decay phase, a break appears in X-ray at $1.6^{+0.7}_{-0.7} \times 10^4$ s and in optical light curve at $7.7^{+1.1}_{-1.1} \times 10^4$ s. Apparently, the X-ray break time is earlier than optical. In fact, this break time is affected by a large uncertainty. In Section 4.1, the X-ray light curve is divided into three segments, i.e., steep decay, shallow decay and late decay, as shown in Table 1. It can also be fit with four-segment broken power law: 1) steep decay (ends at $272.9^{+10.2}_{-9.6}$ s); 2) plateau (from $272.9^{+10.2}_{-9.6}$ s to $1.36^{+0.28}_{-0.90} \times 10^4$ s); 3) normal decay (from $1.36^{+0.28}_{-0.90} \times 10^4$ s to $1.0^{+0.4}_{-0.4} \times 10^5$ s); 4) late decay (starting from $1.0^{+0.4}_{-0.4} \times 10^5$ s) \footnote{\url{https://www.swift.ac.uk/xrt\_live\_cat/00882252/}}. In such case, the X-ray break time ($1.0^{+0.4}_{-0.4} \times 10^5$ s) is roughly consistent with optical  ($7.7^{+1.1}_{-1.1} \times 10^4$ s). Therefore, the X-ray and optical breaking simultaneously cannot be ruled out. 

The forward-shock model predicts temporal indices of $\alpha_O=3(p_{\rm f} - 1)/4$ and $\alpha_X =(3p_{\rm f}-2)/4$ for optical and X-rays at late times, respectively, which correspond to $\alpha_O=0.8$ and $\alpha_X=1.0$ using $p_{\rm f}=2.07$ \citep[see Table 13 therein]{Gao+2013} . Our observed values $\alpha_{O,\rm{late}} = 1.29 \pm 0.06$ and $\alpha_{X,\rm{late}} = 1.31 \pm 0.03$ are larger than those expected. 

\begin{figure*}[htp]
\center
\includegraphics[width=0.9\textwidth]{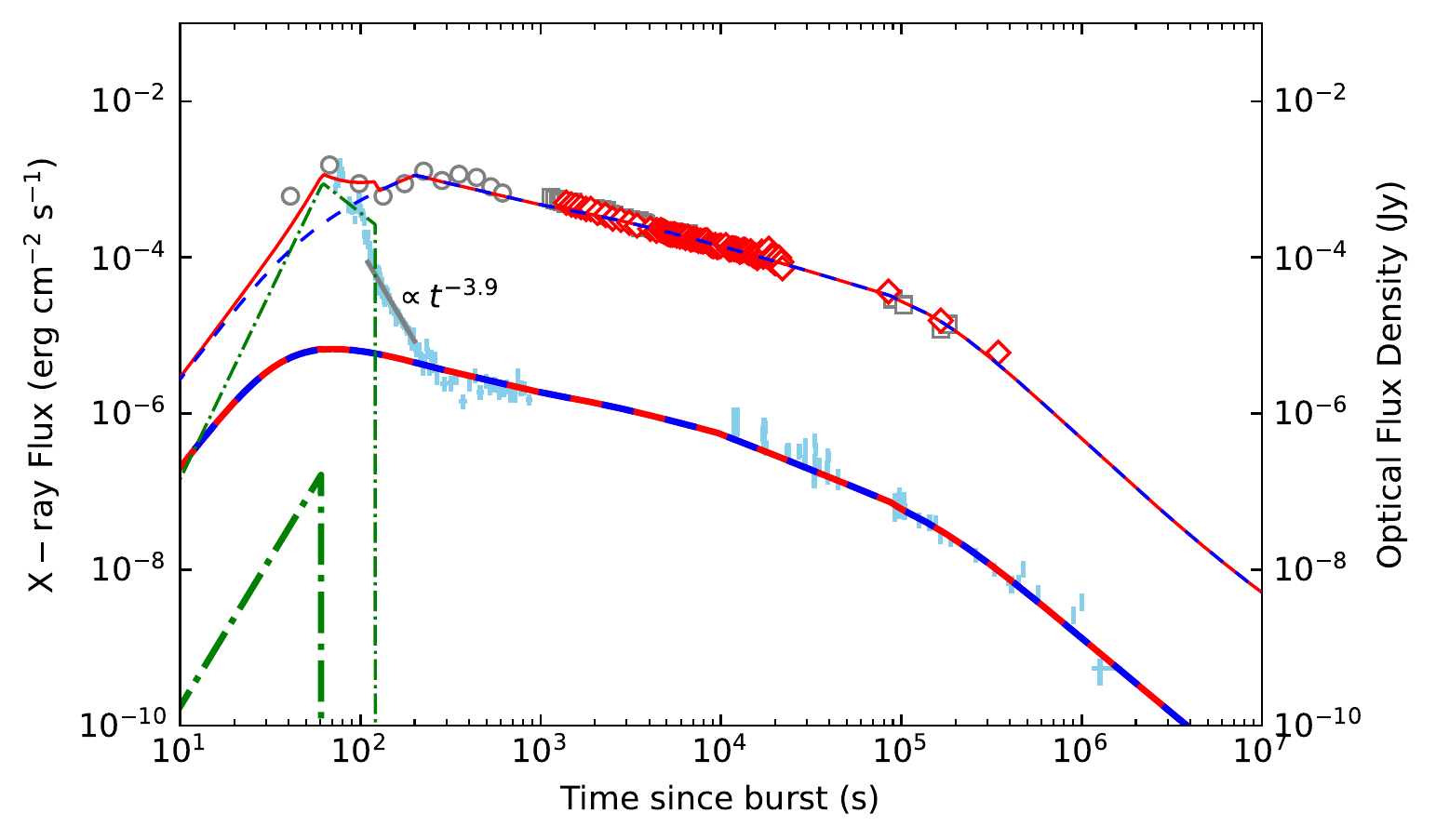}
\caption{The modeling of X-ray (thick lines) and optical \emph{R} band (thin lines) light curves. Observational data are presented with circles for optical (\emph{R} band) and pluses for X-ray. The blue dashed lines represent emission from FS, while green dot-dashed lines for RS. The combined emission from FS and RS are shown with red solid lines. The parameters adopted are listed in Table \ref{tab:parametes}.}
\label{fig:model}
\end{figure*}

Assuming that this break is due to the jet break, the predicted temporal index will become $\alpha_O=1.5$ and $\alpha_X=1.7$ by adopting $p_{\rm f}=2.07$. As shown in Fig.~\ref{fig:model}, our model can roughly fit the late time optical and X-ray light curves. Using this jet break time, we can estimate the opening angle $\theta_{\rm j} \sim 3.3^\circ$ if we inset $E_{\rm K,iso}=9\times 10^{52}$ erg, $n_1=0.1\ \rm cm^{-1}$ and $t_{\rm }\simeq 7\times 10^4/(1+z)$ s into the analytical expression Equation (12). As shown in Table \ref{tab:parametes}, our numerical modeling gives the opening angle $\theta_{\rm j} \sim 4.2^\circ$, which is slightly larger than this analytical estimation.

\begin{table*}[htbp]
	\begin{center}{%\scriptsize
			\caption{Values of parameters adopted for interpreting the broadband data of GRB\,190106A.}
			\label{tab:parametes}
			\begin{tabular}{ccccccc} \hline\hline
				\multicolumn{7}{c}{Forward shock parameters}\\
				\hline
				$E_{\rm K, iso}~({\rm erg})$     & $\Gamma_0$  &  $\theta_{\rm j}~({\rm deg})$ & $n_1~(\rm{cm^{-3}})$ &$\epsilon_{\rm e,f}$ &  $\epsilon_{\rm B,f}$&  $p_{\rm f}$  \\
				$9.0\times 10^{52}$     &  $300$ &  $4.2$   & $0.1$& $0.26$     &  $0.001$&  $2.07$        \\
				\hline
				\multicolumn{7}{c}{Reverse shock parameters}\\
				\hline
				& $\epsilon_{\rm e,r}$ & $\epsilon_{\rm B,r}$  & $p_{\rm r}$ \\
				&$0.002$ &$0.09$     &  $2.1$ \\
				\hline
				
				\hline
				\multicolumn{7}{c}{Energy injection parameters}\\
				\hline
				& $L_{\rm inj}^0 (\rm erg\ s^{-1})$ & $q$  & $t_{\rm start}(1+z)~(\rm{s})$  &$t_{\rm end}(1+z) (\rm s)$ &  \\
				&$0.45 \times 10^{50}$ &$0.05$     &  $300$    & $1500$     &     \\
				\hline\hline
			\end{tabular}
		}
	\end{center}
\end{table*}

\section{Discussion}\label{discussion}

\begin{figure*}
	\centering
	\begin{minipage}{0.49\linewidth}
		\centering
		\includegraphics[width=1\textwidth]{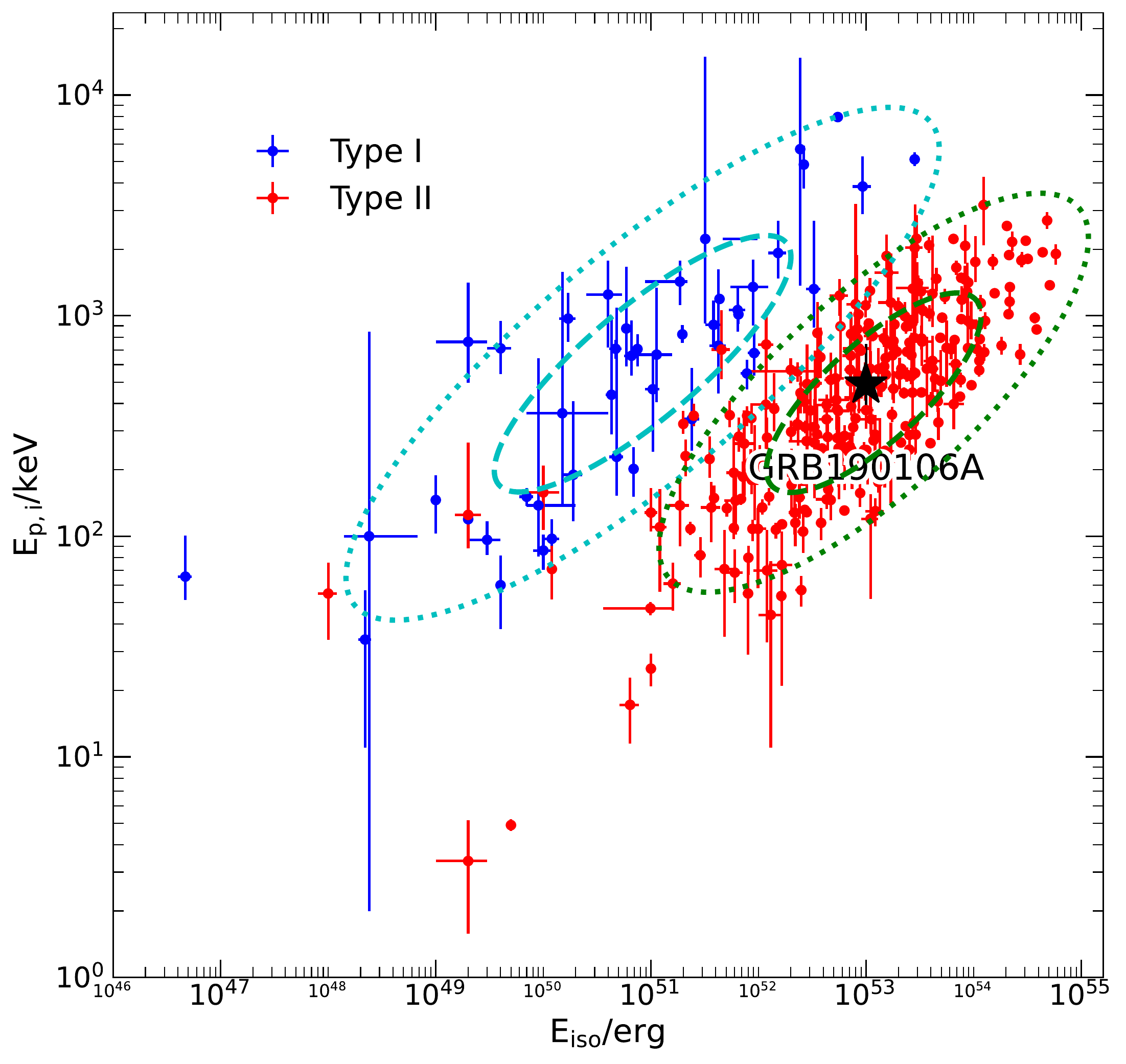}
	\end{minipage}
	\begin{minipage}{0.49\linewidth}
		\centering
		\includegraphics[width=1\textwidth]{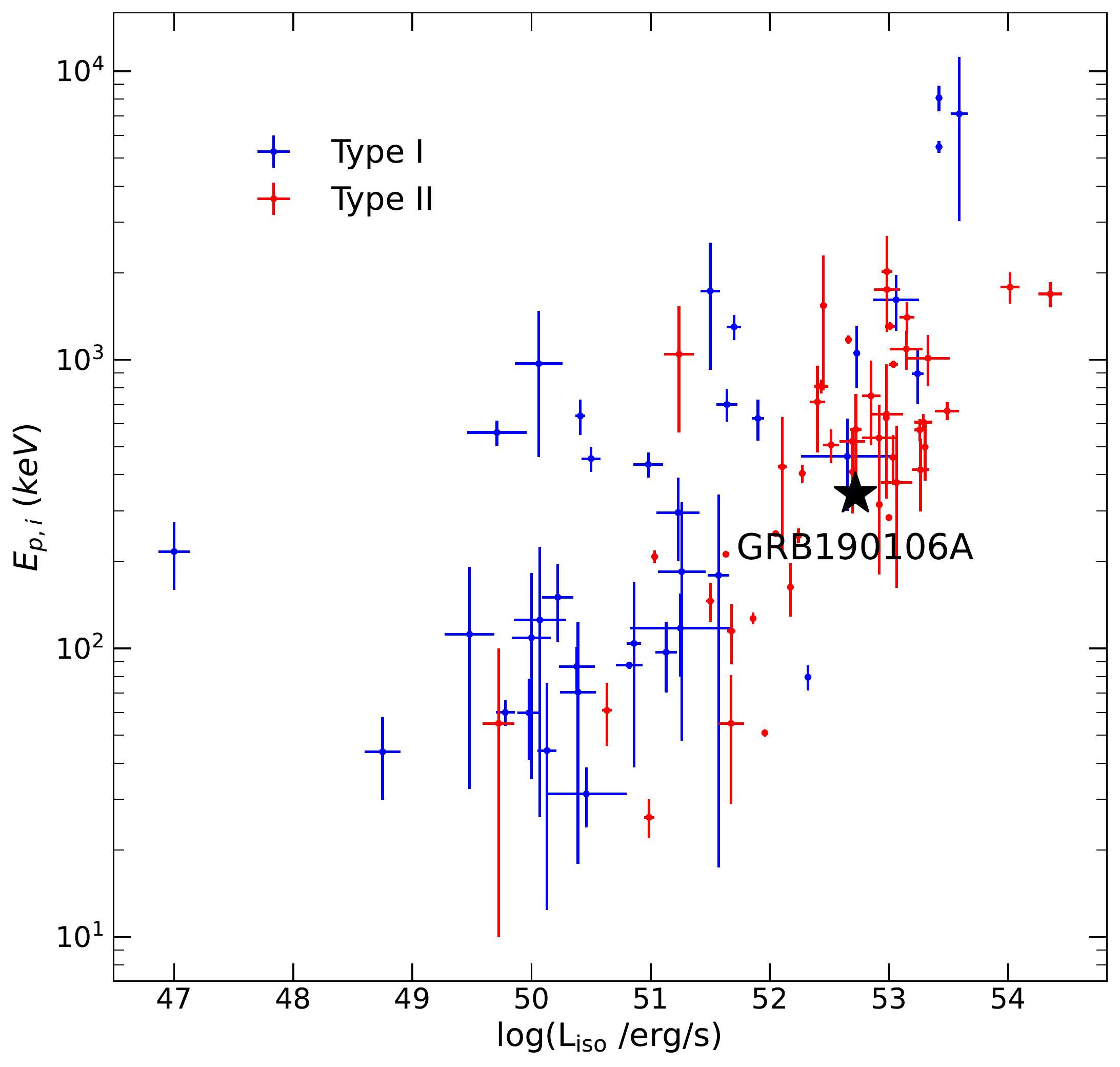}
	\end{minipage}

\caption {\textbf{Left panel:} The Amati diagram of GRB\,190106A (black star). Dashed and dotted lines represent $1\sigma$ and $2\sigma$ confidence levels, respectively. The ellipses colored in cyan and green represent type I and type II burst confidence, respectively. The data points for type I and type II GRBs are taken from \cite{Minaev+2020}. The burst lies within the $1\sigma$ confidence region of type II bursts, consistent with the $E_{p} - E_{iso}$ correlation. \textbf{Right panel:} The Yonetoku diagram of GRB\,190106A (black star). The data points for type I and type II GRBs are taken from \cite{2012MNRAS.421.1256N} and \cite{2022Ap&SS.367...74Z}.} 
\label{fig:amati}
\end{figure*}

With the redshift measurement in this work, we tested the GRB by using two well-known relations, namely the Amati \citep{Amati02,Amati06} and Yonetoku \citep{2004ApJ...609..935Y} relations, as show in Fig.~\ref{fig:amati} left and right panel, respectively. The positions of the burst lie within the $68\%$ prediction regions, consistent with both Amati and Yonetoku relations, which is also mentioned by \citet{Tsvetkova+2019}. The result suggests that GRB\,190106A is a typical type II GRB.

The early double-peak optical light curve was expected by \cite{2003ApJ...595..950Z}. We show that GRB\,190106A is a good case with reverse-forward shocks, just like GRB 041219A \citep{Fan2005} and GRB 110205A \citep{Zheng2012}. Such double-peak (``rebrightening'') light curve is typical for RS+FS emission combinations in a homogeneous ISM case \citep{2003ApJ...595..950Z,Kobayashi2003}. The rising temporal index $\alpha_O=-1.88<-1/2$ favors the RS in thin-shell regime. Therefore, we adopted a thin-shell reverse-forward shock model in a constant ISM case to investigate GRB 190106A. In this scenario, the first optical peak (RS peak) gives the fireball deceleration time \citep{2003ApJ...595..950Z}, which can be used to estimate the initial Lorentz factor of the GRB, $\Gamma_0 \simeq 300$. The second optical peak appears when $\nu_{\rm m}$ crosses the optical band. This peak flux can be used to constrain the values of jet isotropic kinetic energy $E_{\rm K, iso}$, ISM density $n_1$ and the microphysics parameter $\epsilon_{\rm B,f}$. Based on the fit with this scenario, we infer that the magnetic field strength ratio in RS and FS is $B_{\rm r}/B_{\rm f} \sim 9.5$, suggesting a magnetic flux carried by the ejecta from the central engine. Therefore, a magnetized central engine model, i.e., a BH central engine powered by the BZ process or a magnetar central engine, is preferred in GRB 190106A.

It is worth to note that the first optical peak, like GRB\,041219A, can be interpreted with a reverse shock \citep{Fan2005}, internal shocks \citep{Vestrand2005,Wei2007}, or tail emission \citep{Panaitescu2020}. The last two models assume that the prompt optical and $\gamma$-ray emissions have a common origin. However, the radiation mechanism for the prompt $\gamma$-ray emission is still poorly understood \citep{Kumar2008}.

The shallow decays in both X-ray and optical bands demand a nearly constant late-time energy injection with $L_{\rm inj} \simeq 4.5\times 10^{49}\ {\rm erg \ s^{-1}}$ and $q\sim 0.05$ lasting from $\sim 300$ s to $ \sim 1500$ s. We investigated several energy injection sources, such as the spin-down of a magnetar, hyper-accreting BH powered by neutrino annihilation, and BH powered by BZ mechanism. The neutrino annihilation mechanism is less efficient for powering the jet at shallow decay phase. Both magnetar and BH central engine with BZ mechanism can give rise to the energy injection required. In general, the flat light curve feature favors the magnetar central engine model. A BH-accreting central engine can also makes a flat light curve, but requires a very small value of viscosity parameter (and thus a very slow accretion) \citep{Lei2017}. On the other hand, the disk will be dominated by advection and contains a strong wind at his afterglow phase. The light curve will then become steeper due to the suppression by the wind \citep{Lei2017}.

In Section 4.1, we divided the X-ray light curve into three stages (steep decay, shallow decay, late decay). As a result, the break time between shallow decay and late decay in X-ray is significantly different with that in optical band ($1.6^{+0.7}_{-0.7} \times 10^4$ s vs $7.7^{+1.1}_{-1.1} \times 10^4$ s), as show in Table 1. However, the X-ray break time will be roughly consistent with optical if we fit the X-ray light curve with four segments (steep decay, plateau, normal decay, late decay). Considering this uncertainty, we attribute such break to the jet break, revealing an opening angle $\sim 4.2^\circ$. From the observations and modeling, we found that the total jet energy is $E_{\rm total}=E_{\gamma, \rm iso} + E_{\rm K, iso} \simeq 1.9\times 10^{53}$ erg. The opening angle-corrected jet energy will be $E_{\rm j} \sim 5\times 10^{50}$ erg, which is well below the maximum rotational energy of  $3 \times 10^{52}$ erg \citep{2016PhR...621..127L} $-$ $7 \times 10^{52}$ erg \citep{2009A&A...502..605H} for a standard neutron star with mass $M\sim1.4\,M_\sun$. Therefore, our data do not require a BH as the central engine of this GRB.

\section{summary}\label{conclusion}

We present our optical observation of the afterglow of GRB 190106A with the NEXT telescope, and the Xinglong 2.16-m telescope equipped with BFOSC. With the spectrum obtained from BFOSC, we measured the redshift of the burst and calculated the Line Strength Parameter. The unusual optical light curve obtained by NEXT and Xinglong 2.16-m, combined with the MASTER observations, show a normal decay light curve with early double peak and late break. Therefore, the optical light curve can be divided into five stages: rapid rise in the early stage, followed by decay, then a slow rise reaching a second peak at around 200 s, followed by a shallow decay (plateau), and faster decay from about 1 d onward. The evolution of the X-ray light curve can be divided more simply into three stages: rapid decay, plateau and normal decay.

The multi-band observations, especially the early optical data of GRB 190106A, provide rich information to study the nature of GRB. It is found that GRB 190101A can be successfully explained by the reverse-forward shock model. Our conclusions are summarized as follows:

1. The redshift of the burst is $z= 1.861\pm 0.002$. The Line Strength Parameter, which is very similar to the average value of the sample.

2. The studies with the Amati and Yonetoku diagrams indicate that GRB\,190106A is a typical type II burst.

3. The early double-peak optical light curve can be well interpreted with combination of the RS and FS models.

4. The modeling with RS and FS models indicate that the magnetic field strength ratio in RS and FS is $B_{\rm r}/B_{\rm f} \sim 9.5$, suggesting a magnetic flux carried by the ejecta from the central engine. This favors a strongly magnetized central engine model, such as a magnetar central engine model or a BH central engine powered by the BZ process.

5. The initial Lorentz factor can be estimated with the first optical peak time (or deceleration time) as $\Gamma_0 \sim 300$. 

6. The multi-band afterglow data can be interpreted with external shock model. From the observations and modeling, we found that the total jet energy is $E_{\rm total}=E_{\gamma, \rm iso} + E_{\rm K, iso} \simeq 1.9\times 10^{53}$ erg.

7. The shallow decays in both X-ray and optical bands are explained by nearly constant late-time energy injection with $L_{\rm inj} \simeq 4.5\times 10^{49}\ {\rm erg \ s^{-1}}$ and $q\sim 0.05$ lasting for $\sim 1000$ s. Both magnetar and BH central engine models can give rise to such amount of injected energy, but the flat light curve feature favors a magnetar central engine.

8. The breaks at a few $\times 10^4$ s in both X-ray and optical bands are roughly consistent with the jet break, revealing an opening angle $\sim 4.2^\circ$. The opening angle-corrected jet energy will be $E_{\rm j} \sim 5\times 10^{50}$ erg.

\section*{acknowledgments}
We are very grateful to Yong Ma for the help with the NEXT observations. We acknowledge the support of the staff of the Xinglong 2.16-m telescope. This work was also partially supported by the Open Project Program of the Key Laboratory of Optical Astronomy, National Astronomical Observatories, Chinese Academy of Sciences. This work is supported by the National Key R\&D Program of China (Nos. 2020YFC2201400), the National Natural Science Foundation of China under grants U2038107, U1931203 and 12021003. D.X. acknowledges support by the science research grants from the China Manned Space Project with NO. CMS-CSST-2021-A13 and CMS-CSST-2021-B11. W. H. Lei and H. Gao acknowledge support by the science research grants from the China Manned Space Project with NO.CMS-CSST-2021-B11. W. Xie acknowledges support from the Guizhou Provincial Science and Technology Foundation (Grant No. QKHJC-ZK[2021] 027). H. Gao acknowledges support from the National SKA Program of China (grant No. 2022SKA0130101). J. Z. Liu acknowledges the science
research grants from the China Manned Space Project with
No. CMSCSST-2021-A08. Data resources are supported by the China National Astronomical Data Center (NADC) and the Chinese Virtual Observatory (China-VO). This work is supported by the Astronomical Big Data Joint Research Center, co-founded by the National Astronomical Observatories, the Chinese Academy of Sciences, and the Alibaba Cloud. We acknowledge the use of public data from the {\em Swift} data archive.

\software{Astrometry.net \citep{Lang+2010},
            Astropy \citep{2022ApJ...935..167A},
            IRAF \citep{Tody+1986},
            Matplotlib \citep{Hunter:2007},
            NumPy \citep{harris2020array},
            Python \citep{10.5555/1593511},
            SciPy \citep{2020SciPy-NMeth},
          Source Extractor \citep{BertinArnouts+1996}
          }

\bibliography{msNotes}{}
\bibliographystyle{aasjournal}

\newpage
\begin{table*}[htbp]
\centering
\caption{The photometric results of NEXT and Xinglong 2.16-m. $\rm T_{mid}$ is the middle time of the exposure after the BAT trigger. Exp is the exposure time. All data are calibrated using nearby {\em SDSS} reference stars and \emph{not} corrected for Galactic extinction, which is  $E(B-V)=0.08$ mag \citep{Schlafly11}.}
\label{table:light curve}
\begin{tabular}{ c c c c c | c c c c c }
\hline 
\hline 
$\rm T_{mid}$ (s) & Exp (s) & Filter & Mag (Vega) & Telescope  
& $\rm T_{mid}$ (s) & Exp (s) & Filter & Mag (Vega) & Telescope  \\ 
\hline
1387&60&$R$&16.92 +/- 0.02&NEXT&	10661&200&$R$&18.27 +/- 0.04&NEXT\\
1482&60&$R$&16.98 +/- 0.02&NEXT&	11109&200&$R$&18.39 +/- 0.05&NEXT\\
1569&60&$R$&17.01 +/- 0.02&NEXT&	11337&200&$R$&18.41 +/- 0.05&NEXT\\
1673&90&$R$&17.05 +/- 0.02&NEXT&	11564&200&$R$&18.36 +/- 0.05&NEXT\\
1786&90&$R$&17.10 +/- 0.02&NEXT&	11791&200&$R$&18.40 +/- 0.05&NEXT\\
1898&90&$R$&17.11 +/- 0.02&NEXT&	12017&200&$R$&18.39 +/- 0.05&NEXT\\
2068&200&$R$&17.24 +/- 0.01&NEXT&	12239&200&$R$&18.42 +/- 0.05&NEXT\\
2291&200&$R$&17.30 +/- 0.02&NEXT&	12462&200&$R$&18.42 +/- 0.05&NEXT\\
2513&200&$R$&17.42 +/- 0.02&NEXT&	12733&300&$R$&18.47 +/- 0.04&NEXT\\
2786&300&$R$&17.45 +/- 0.02&NEXT&	13052&300&$R$&18.45 +/- 0.04&NEXT\\
3105&300&$R$&17.53 +/- 0.02&NEXT&	13370&300&$R$&18.41 +/- 0.05&NEXT\\
3425&300&$R$&17.64 +/- 0.02&NEXT&	13688&300&$R$&18.43 +/- 0.05&NEXT\\
4063&300&$R$&17.76 +/- 0.02&NEXT&	14005&300&$R$&18.50 +/- 0.05&NEXT\\
4388&300&$R$&17.80 +/- 0.02&NEXT&	14321&300&$R$&18.52 +/- 0.05&NEXT\\
4654&120&$R$&17.77 +/- 0.03&NEXT&	14637&300&$R$&18.47 +/- 0.04&NEXT\\
4799&120&$R$&17.84 +/- 0.03&NEXT&	14955&300&$R$&18.55 +/- 0.05&NEXT\\
4945&120&$R$&17.86 +/- 0.03&NEXT&	15272&300&$R$&18.62 +/- 0.05&NEXT\\
5089&120&$R$&17.90 +/- 0.03&NEXT&	15589&300&$R$&18.61 +/- 0.06&NEXT\\
5236&120&$R$&17.94 +/- 0.04&NEXT&	15905&300&$R$&18.69 +/- 0.06&NEXT\\
5381&120&$R$&17.90 +/- 0.03&NEXT&	16222&300&$R$&18.63 +/- 0.05&NEXT\\
5529&120&$R$&17.92 +/- 0.03&NEXT&	16539&300&$R$&18.54 +/- 0.05&NEXT\\
5675&120&$R$&17.98 +/- 0.04&NEXT&	16855&300&$R$&18.48 +/- 0.05&NEXT\\
5912&200&$R$&18.00 +/- 0.03&NEXT&	17172&300&$R$&18.62 +/- 0.06&NEXT\\
6138&200&$R$&17.97 +/- 0.03&NEXT&	17489&300&$R$&18.52 +/- 0.05&NEXT\\
6363&200&$R$&18.04 +/- 0.03&NEXT&	17806&300&$R$&18.59 +/- 0.06&NEXT\\
6589&200&$R$&18.01 +/- 0.03&NEXT&	18123&300&$R$&18.61 +/- 0.07&NEXT\\
6813&200&$R$&18.09 +/- 0.03&NEXT&	18440&300&$R$&18.41 +/- 0.06&NEXT\\
7042&200&$R$&18.05 +/- 0.03&NEXT&	18756&300&$R$&18.68 +/- 0.07&NEXT\\
7267&200&$R$&18.10 +/- 0.03&NEXT&	19074&300&$R$&18.64 +/- 0.07&NEXT\\
7491&200&$R$&18.14 +/- 0.03&NEXT&	19391&300&$R$&18.63 +/- 0.07&NEXT\\
7717&200&$R$&18.12 +/- 0.03&NEXT&	19709&300&$R$&18.84 +/- 0.09&NEXT\\
7946&200&$R$&18.16 +/- 0.03&NEXT&	20025&300&$R$&18.83 +/- 0.10&NEXT\\
8172&200&$R$&18.12 +/- 0.03&NEXT&	20342&300&$R$&18.85 +/- 0.10&NEXT\\
8397&200&$R$&18.12 +/- 0.03&NEXT&	20658&300&$R$&18.75 +/- 0.10&NEXT\\
8624&200&$R$&18.23 +/- 0.04&NEXT&	20975&300&$R$&18.72 +/- 0.10&NEXT\\
8852&200&$R$&18.28 +/- 0.04&NEXT&	21292&300&$R$&18.96 +/- 0.13&NEXT\\
9078&200&$R$&18.30 +/- 0.04&NEXT&	21609&300&$R$&18.84 +/- 0.12&NEXT\\
9302&200&$R$&18.28 +/- 0.04&NEXT&	21925&$5\times300$&$R$&19.12 +/- 0.12&NEXT\\
9529&200&$R$&18.32 +/- 0.04&NEXT&	84911&$5\times360$&$R$&19.77 +/- 0.06&Xinglong 2.16-m\\
9756&200&$R$&18.30 +/- 0.04&NEXT&	165878&$7\times360$&$R$&20.70 +/- 0.05&Xinglong 2.16-m\\
9980&200&$R$&18.31 +/- 0.04&NEXT&	345290&$10\times360$&$R$&21.73 +/- 0.06&Xinglong 2.16-m\\
\hline 
\hline
\end{tabular}
\end{table*}

\newpage

\begin{table*}[htbp]
	\begin{center}{%\scriptsize
			\caption{List of features in the spectra and their observatory frame equivalent widths.}
			\label{table:spectral lines}
			\begin{tabular}{cccc} \hline\hline
				$\rm \lambda_{obs}$(\AA) & Feature(\AA) & $z$ & $\rm EW_{obs}$(\AA) \\
				\hline
4428.26 & \ion{ C }{4}/\ion{ C }{4}$\lambda\lambda$1549 & 1.859 & $8.68 \pm 1.04 $ \\
4604.72 & \ion{ Fe }{2}$\lambda\lambda$1608 & 1.864 & $2.11 \pm 0.63 $ \\
4776.51 & \ion{ Al }{2}$\lambda\lambda$1670 & 1.860 & $4.84 \pm 0.55 $ \\
5304.92 & \ion{ Al }{3}$\lambda\lambda$1854 & 1.861 & $3.30 \pm 0.49 $ \\
5329.60 & \ion{ Al }{3}$\lambda\lambda$1862 & 1.862 & $1.44 \pm 0.35 $ \\
6704.71 & \ion{ Fe }{2}$\lambda\lambda$2344 & 1.860 & $4.25 \pm 0.41 $ \\
6789.60 & \ion{ Fe }{2}$\lambda\lambda$2374 & 1.860 & $3.01 \pm 0.45 $ \\
6811.21 & \ion{ Fe }{2}$\lambda\lambda$2382 & 1.859 & $7.52 \pm 0.45 $ \\
7397.92 & \ion{ Fe }{2}$\lambda\lambda$2586 & 1.861 & $3.59 \pm 0.55 $ \\
7435.47 & \ion{ Fe }{2}$\lambda\lambda$2600 & 1.860 & $5.45 \pm 0.64 $ \\
8011.58 & \ion{ Mg }{2}/\ion{Mg}{2}$\lambda\lambda$  2800 & 1.861 & $14.14 \pm 1.01 $ \\
8158.45 & \ion{ Mg }{1}$\lambda\lambda$  2852 & 1.861 & $2.23 \pm 0.44 $ \\
				\hline\hline
			\end{tabular}
		}
	\end{center}
\end{table*}

\end{document}